\documentclass[fleqn,usenatbib]{mnras}
\usepackage{newtxtext,newtxmath}
\usepackage[T1]{fontenc}
\usepackage{ae,aecompl}
\usepackage{color,soul}

\usepackage{graphicx}	% Including figure files
\usepackage{amsmath}	% Advanced maths commands
\usepackage{amssymb}	% Extra maths symbols

\title[Faint  Low-Frequency Radio Source Population]{The Nature of the Faint Low-Frequency Radio Source Population}

\author[Ocran et al.]{
E.\ F. Ocran$^{1,2}$, A.\ R.\ Taylor$^{1,2}$, M.\ Vaccari$^{1,3}$, D.\ A.\ Green$^4$, 
\\
$^1$ Department of Physics and Astronomy, University of the Western Cape, Private Bag X17, Bellville 7535, South Africa \\
$^2$ Department of Astronomy, University of Cape Town, Private Bag X3, Rondebosch 7701, South Africa \\
$^3$ INAF - Istituto di Radioastronomia, via Gobetti 101, 40129 Bologna, Italy \\
$^4$ Astrophysics Group, Cavendish Laboratory, Cambridge University, 19 J.\ J.\ Thompson Ave., Cambridge, CB3 0HE, UK
}

\date{Accepted 2017 February 16. Received 2017 February 15; in original form 2016 December 17}

\pubyear{2017}

\begin{document}
\label{firstpage}
\pagerange{\pageref{firstpage}--\pageref{lastpage}}
\maketitle

\begin{abstract}
We present a multi-wavelength study into the nature of faint radio sources in a deep radio image with the Giant Meterwave Radio Telescope at 612\,MHz covering 1.2 deg$^2$ of the ELAIS N1 region.  
We detect 2800 sources above 50~$\mu$Jy\,beam$^{-1}$.  By matching to multi-wavelength data we obtain a redshift estimate for 63\%, with 29\% based on spectroscopy. 
For 1526 of the sources with redshifts we use radio and X-ray luminosity, optical spectroscopy, mid-infrared colors, and 24$\mu$m and IR to radio flux ratios to search for the presence of an AGN.
The analysis reveals a rapid change in the population as flux density decreases from 
$\sim$500\,$\mu$Jy to $\sim$100\,$\mu$Jy.
We find that 80.3\% of the objects show no evidence of AGN and have multi-wavelength properties consistent with radio emission from star forming galaxies (SFG). 
We classify 11.4\% as Radio Quiet (RQ) AGN and the remaining 8.3\% as Radio Loud (RL) AGN.

The redshift of all populations extends to $z > 3$ with a median of $\sim$1. The median radio and far-IR luminosity increases systematically from SFG, to RQ AGN and RL AGN.  
The median $q_{\rm 24 \mu m}$ for SFG, $0.89\pm0.01$ is slightly below that for RQ AGN, $1.05\pm0.03$, and both differ substantially from the value for RL AGN of $-0.06\pm0.07$. 
However, SFG and RQ AGN show no significant difference in far-IR/radio ratios and have statistically indistinguishable star formation rates inferred from radio and far-IR luminosities.   
We conclude that radio emission from host galaxies of RQ AGN in this flux density regime result primarily from star formation activity.

\end{abstract}

\begin{keywords}
galaxies: active, infrared: galaxies, radio continuum: galaxies
\end{keywords}

%%%%%%%%%%%%%%%%%%%%%%%%%%%%%%%%%%%%%%%%%%%%%%%%%%

%%%%%%%%%%%%%%%%% BODY OF PAPER %%%%%%%%%%%%%%%%%%

\section{Introduction}

The interplay between star formation and black hole accretion is one of the central issues in galaxy evolution today. 
Tight correlations found between black hole mass and both stellar bulge mass \citep{1998AJ....115.1823K, 2002MNRAS.331..795M} 
and velocity dispersion \citep{2000ApJ...539L..13G, 2002ApJ...574..740T} in galaxies in the local universe indicates that the 
formation and growth of galaxies is closely linked to the growth of their central black holes.

Both accretion activity on the central black holes and star formation in the disks of galaxies create radio synchrotron emission. 
In this context, the study of the faint radio universe has become a very active field of research as the increased
 sensitivity of radio telescopes has enabled $\mu$Jy-rms-level imaging over extended fields \citep{2015AJ....150...87L},
 probing radio emission from galaxies to significant red shift. 
 Understanding the properties of sources in the faint radio sky and their cosmic evolution is thus important 
 to understanding of the links between black hole accretion and star formation activity in the universe 
 \citep{2009MNRAS.395.1249C}. 
 
Whereas the bright (i.e. $>1$ mJy at 1.4 GHz) radio sky is dominated by the emission driven by `radio-loud' AGNs 
(RL AGN) (e.g. \citealt{2001A&A...365..392P, 2008MNRAS.386.1695S, 2009ApJ...694..235P} ), deep radio surveys 
probe both active galactic nuclei (AGN) and star-forming galaxies (SFG), with SFGs becoming increasingly important at fainter flux densities. 
Recent work by \cite{2009ApJ...694..235P} and \cite{2013MNRAS.436.3759B} has revealed a third population of sources 
at faint flux densities, the `radio-quiet' AGN (RQ AGN). \cite{2009ApJ...706L.260G} proposed that RQ AGN represent 
scaled-down versions of RL AGN in mini radio jets. On the other hand, \cite{2011ApJ...739L..29K} and \cite{2011ApJ...740...20P} 
argued that the radio emission of RQ AGN come from star formation in the host galaxy.

Radio flux density measurements at angular resolutions insufficient to resolve the structure of the galaxy
cannot distinguish between synchrotron emission from AGN or star formation.
However, multi-wavelength observations from the X-ray to the millimetre can reveal the presence of AGN, including 
highly obscured ones. 
\cite{2015MNRAS.452.4111R} divided faint radio samples into sources powered by AGN or by star formation through 
their infrared SEDs, X-ray luminosities and IRAC colors satisfying the \cite{2012ApJ...748..142D} criterion. 
\cite{2013MNRAS.436.3759B} proposed a classification scheme which is an upgrade of that used by
 \cite{2011ApJ...740...20P} through a combination of radio, infrared and X-ray data in the ECDFS.

In this work, we present a multi-wavelength investigation to classify radio sources from deep 612 MHz observations 
with the GMRT of the ELAIS N1 field covering an area of 1.2 deg$^2$ to a RMS sensitivity of  10\,$\mu$Jy\,beam$^{-1}$.
These observations are the most sensitive images of the sky at this frequency to date.
\cite{2008MNRAS38375G} presented a GMRT image of ELAIS N1 at 610 MHz covering $\sim$9\,deg$^2$, constructed  
from a mosaic of 19 pointings.  
This comprised a smaller central region of 4 pointings with an rms of $\sim$40\,$\mu$Jy\,beam$^{-1}$ and the remaining 
15 pointings with an rms of $\sim$70\,$\mu$Jy\,beam$^{-1}$. 
\cite{2010ApJ...714.1689G} imaged 15\,deg$^2$ at 1.4 GHz with the Dominion Radio Astrophysical Observatory 
to an rms of $55\,\mu$Jy\,beam$^{-1}$ in total intensity and  $45\,\mu$Jy\,beam$^{-1}$ in linear polarization.
 While stacking analysis of such deep radio images based on optical-IR catalogues has allowed statistical 
 correlations such as the average radio-IR correlation to be explored to lower radio flux densities 
 \citep{2009MNRAS.394..105G}, direct studies of the radio selected population to $\mu$Jy sensitives, 
 ultra-deep radio imaging. 
  
The paper is organized as follows.  We describe the radio observations and the multi-wavelength data and sample selection
in  Section~\ref{multi.sec}.  The multi-wavelength AGN diagnostics are described  in Section~\ref{agn.sec}, 
and the properties of the classified sources in Section~\ref{results.sec}. We summarize our results in Section~\ref{sum.sec}. 
For calculation of intrinsic source properties we assume a flat cold dark matter ($\rm{\Lambda}$CDM) 
cosmology with  $\rm{\Omega_{\Lambda} \ = \ 0.7}$, $\rm{\Omega_{m} \ = \ 0.3}$ and $\rm{H_{o} \ = \ 70 \ km\,s^{-1} \ Mpc^{-1}}$ .

\section{Multi-Wavelength Data}\label{multi.sec}
 
\subsection{Radio Data}\label{radiodata.sec}
 
GMRT observations of the ELAIS N1 field were obtained during several observing runs from 2011 to 2013. 
Observations were carried out for 7 positions arranged in a hexagonal pattern centred on 
$\mathrm{\alpha \ = \ 16^{h} \ 10^{m} \ 30^{s}, \  \delta \ = \ 54^{\circ} \ 35 \ 00^{\prime \prime}}$, covering an 
area of 1.2~deg$^2$. Each position was observed for $\rm{\sim}$30 hours in three 10-hour sessions. 
The total bandwidth was 32 MHz, split into 256 spectral channels centered at 612 MHz in four polarization states. 
Observations of 3C286 were made twice in each observing session, and
were used to calibrate the flux scale, band pass and absolute 
polarization position angle. Time dependent gains and on-axis polarisation leakage corrections were measured by frequent observations of 
the secondary calibrator J1549+506. 

The visibility data were calibrated, imaged and mosaicked using the CASA processing software. 
Images were constructed for each pointing out to the 10\% point of the primary beam 
using multi-frequency synthesis with two Taylor series terms.  
The effect of the 3-dimensional beam over the wide field-of-view was corrected using w-projection.
Following initial calibration, the visibilities for each pointing were run through several iterations of self-calibration, 
with the gain solution time interval decreasing and the depth of the 
clean component sky model increasing with each iteration.
The self-calibrated individual field images were combined into a mosaic in 
the image plane using the CASA linearmosaic tool.
The RMS noise in the resulting total intensity mosaic image is 10.3\,$\mu$Jy\,beam$^{-1}$ before primary beam correction.  
 The RMS was measured by fitting a normal error function to 
distribution of negative map deflections from the entire mosaic. 
By comparison, the RMS in the Stokes $Q$ and $U$ images 
is 7.3\,$\mu$Jy\,beam$^{-1}$. 
The RMS noise in the total intensity mosaic is thus only slightly above the thermal noise limit. 
The excess noise may be attributed to a combination of confusion and residual clean artifacts.

\begin{figure}
  \includegraphics[width = \columnwidth]{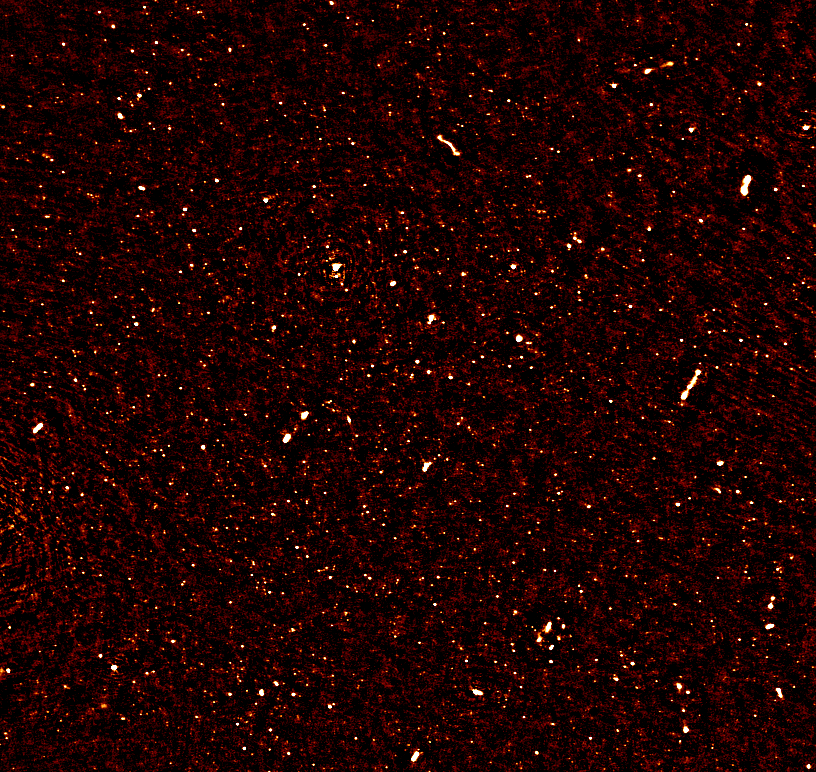}
  \caption{ Image of the ELAIS\,N1 GMRT 612 MHz mosaic
  centered at 
  $\mathrm{\alpha \ = \ 16^{h} \ 10^{m} \ 30^{s}, \  \delta \ = \ 54^{\circ} \ 35 \ 00^{\prime \prime}}$.  
 This image is 47$'$ on each side, showing the central $\sim$0.6 deg$^2$ of the 1.2 deg$^2$ mosaic.
 The RMS noise is 10.3\,$\mu$Jy\,beam$^{-1}$.  
 The image shows a small number of bright classical radio galaxies with double-lobed and jet 
  morphologies.  However, the vast majority of the radio sources are fainter objects that are unresolved at the $\sim5''$
  angular resolution. }
  \label{fig_gmrtimage} 
\end{figure}

 The angular resolution of the mosaic radio image is $6.1\times5.1$\,arcsec.  An image of the central 0.6 deg$^2$ of the mosaic is shown in Figure~\ref{fig_gmrtimage}.
 
 The catalogue of radio sources was extracted using the AIPS SAD automated source finding program
 and consists of 2800 radio sources with peak flux density above the $5\sigma$ threshold. 
 Among the brightest objects in the sample, is the very small fraction of the total ($\sim$3\%) 
 that exhibit classical extended radio-galaxy, AGN-driven jet morphology 
 \citep{2016MNRAS.459L..36T}.  
 The vast majority are compact sources that are unresolved at the resolution of the observations. 
The flux densities for compact individual sources was derived from the 2-D Gaussian fits
from AIPS SAD.  Total flux densities for the small number of extended multi-component jet-lobe 
sources was taken  to be the sum of the flux densities of the individual components.
The full catalogue, including source counts and polarization information,
will be published  in a subsequent paper.  Here we focus on matching with
the multi-wavelength data sets to investigate the nature of the 
radio sources.

% %---------------------------------------------------------------------------------------------------------------------------
 \subsection{SERVS Data Fusion and Ancillary Data}\label{df.sec}
 \indent
The Spitzer Extragalactic Representative Volume Survey (SERVS, \citealt{Mauduit2012}) is a warm Spitzer survey 
which imaged $\rm{18 \ deg^{2}}$ using the IRAC1 3.6 $\rm{\mu m}$ and IRAC2 4.5 $\rm{\mu m}$ bands. The main 
aim of the survey was to enable the study of galaxy evolution as a function of environment from $\rm{z \sim 5}$ to the 
present day through the first extragalactic survey that is both large enough and deep enough to put rare objects such 
as luminous quasars and galaxy clusters at ${z \ge 1}$ into their cosmological context.  SERVS sources were separately 
extracted in IRAC1 and IRAC2 images and single-band detections were merged into a two-band (hereafter IRAC12) 
catalogue using a search radius of 1 arcsec. The rms noise of both the IRAC1 and IRAC2 images is $\rm{0.5\,\mu Jy}$ 
and the catalogue completeness limit is thus $\rm{2\,\mu Jy}$ in both bands.

SERVS overlaps with several other surveys at optical, near- through far-infrared, sub-millimeter. The SERVS Data Fusion\footnote{\url{http://www.mattiavaccari.net/df}} (\citealt{Vaccari2010,Vaccari2015}), matches SERVS IRAC12 sources against a large suite of multi-wavelength data using a search radius of 1 arcsec. In particular, in our work we have used the following datasets included within the SERVS Data Fusion:

\begin{itemize}
\item SDSS DR12 Optical Photometry \citep{Alam2015}
\item INT WFC Deep Optical Photometry \citep{GonzalezSolares2011}
\item 2MASS PSC Near-IR Photometry \citep{Cutri2003}
\item UKIDSS DXS Deep Near-IR Photometry \citep{Lawrence2007}
\item Spitzer Wide-Area Infrared Survey 4-band IRAC Photometry and 3-band MIPS Photometry \citep{Lonsdale2003,Vaccari2015} 
\item SWIRE Photometric Redshifts \citep{RowanRobinson2008,RowanRobinson2013,RowanRobinson2016}
\item Spectroscopic redshifts from the literature \citep{Serjeant2004,Trichas2009,Hernan-Caballero2009,Alam2015} 
including newly obtained measurements from BOSS/SDSS-III better detailed in Sec.~\ref{bossdata.sec} 
\end{itemize}

In future papers, we will improve our analysis by making use of the multi-wavelength database produced by the Herschel Extragalactic Legacy Project \citep[HELP]{Vaccari2016}, including an improved source extraction at SPIRE \cite{Griffin2010} wavelengths using the XID+ software by \cite{Hurley2017}.
% %---------------------------------------------------------------------------------------------------------------------------

\subsection{BOSS/SDSS Spectroscopy}\label{bossdata.sec}
 
The Baryon Oscillation Spectroscopic Survey (BOSS) was the primary dark-time survey of SDSS-III \cite{Eisenstein2011}, 
the third phase of the Sloan Digital Sky Survey (SDSS, \citealt{2000AJ....120.1579Y}). BOSS observed 1.5 million massive galaxies and 150,000 quasars to measure the distance-redshift relation ${d_{A} (z)}$ and the Hubble parameter $H(z)$ with percent-level precision out to $z = 0.7$ and $z \simeq 2.5$, using the well-established techniques that led to the first detection of the baryon acoustic oscillations (BAO) feature  \citep{2005MNRAS.362..505C,2005ApJ...633..560E}. 
BOSS was designed to measure the scale of the BAO in the clustering of matter over a larger volume than the combined 
efforts of all previous spectroscopic surveys of large-scale structure. BOSS consists of two spectroscopic surveys, both of which have been carried out over an area of 10,000 deg$^{2}$. The first survey observed 1.5 million luminous red galaxies ($i < 19.9$) to measure BAO to redshifts $z < 0.7$, while the second one will observe neutral hydrogen in the Ly$\alpha$ forest in more than 150,000 quasar spectra ($g < 22$) to constrain BAO at $z \sim 2.5$.

From 2012 onwards, a series of plates were added to the SDSS-III program to observe ancillary science programs. We were granted four BOSS plates to obtain spectra for the radio sources detected by our GMRT observations as 
well as by other shallower radio surveys in the ELAIS N1 field. Observations were obtained by the SDSS team and their reduction using the default BOSS spectroscopic pipeline has been made publicly available as part of the SDSS DR12. In our work, however, due to the large number of sources near the center of the BOSS plate, a bespoke data reduction scheme was adopted to optimize the spectroscopic redshift accuracy (Tarr et.\ al.\ in prep).

\subsection{X-Ray Data}\label{xraydata.sec}

The only astrophysical sources that reach very high X-ray luminosities are AGNs. Usually a cut in the unabsorbed 
X-ray luminosity at 10$^{42}$ erg s$^{-1}$ is considered as a threshold dividing AGNs and SFGs 
(see \citealt{2004ApJS..155..271S}).
In this work we combined shallow Chandra observations of the whole GMRT field produced by \cite{Trichas2009} 
using the Imperial College London pipeline (\citealt{2008MNRAS.388.1205G,2009ApJS..180..102L}) with deep 
Chandra observations of a small area (0.15 deg$^2$) of the central region of the field by \cite{Manners2003}.

\section{Multi-Wavelength Cross-Matching}\label{matching.sec}

We matched the GMRT catalogue against SERVS IRAC12 positions using a variable search radius equal to three 
times the estimated astrometric error. 
Radio position errors for individual sources from AIPS SAD fitting algorithm are
typically a few a tenths of an arcsecond. SERVS astrometric error of 0.15 arcsec was assumed based on SERVS 
vs 2MASS cross-matching.  
From an initial matching of the GMRT and SERVS catalogues we computed the median astrometric offsets between the two catalogues of $-0.434 \pm 0.014$ and $+0.303 \pm 0.013$ arcsec 
in RA and DEC respectively.
We applied these correction to the radio positions before performing the final cross-matching. SERVS matches were identified using a search radius of three times the combined astrometric
error on the coordinate comparison.  Given the sub-arcsecond accuracy of the positions,in 
virtually all cases where a match was found this resulted in a unique identification.  In the few
cases of a multiple SERVS objects in error circle we took the nearest object as the
identification.

SERVS identifications were obtained for 2369 out of 2800 (or 85\%) of our radio sources. The RMS of the coordinate differences for the matched sources is $\sim$0.5$''$ in both
RA and DEC.
Of the 15\% of GMRT objects that are not matched in the SERVS catalogue, 203 (7.3\%) are affected by diffraction spikes due to bright objects and/or by source blending due to confusion, and therefore no conclusive identifications can be made. This can partly be addressed by carrying out multi-band forced photometry with e.g. the Tractor code \citep{Lang2016}, as currently underway within the SERVS team (Nyland et al. in prep). Another 109 objects (3.9\%) have faint counterparts  in the SERVS images but have no entry in the SERVS catalogue.  
The remaining 124 unidentified objects (4.4\%) have no infrared counterpart in the SERVS image.  These objects are examples of the recently identified class of Infra-red Faint Radio Sources \citep{2006AJ....132.2409N,Garn:2008cx}.

Table~\ref{tab_matches} summarizes the results of the cross-matching against SERVS and the other multi-wavelength data sets. The fraction of identifications in either SERVS IRAC1 or IRAC2 is 85$\%$, while 80$\%$ have counterparts in both IRAC1 and IRAC2 and 39\% of our radio sample has counterparts in all four IRAC bands. MIPS~24 detections are found for 60\%. In addition, $3.3\%$ of our sample has an X-ray detection in either the shallow or deep X-ray ($2.5\%$ and $0.8\%$ respectively).

 \begin{table}
 \centering
 \caption{GMRT cross-matching statistics. The table shows statistics from our new cross-matching.}
 \begin{tabular}{lrr}
 \hline
 \hline
 Catalogue    & Size & Fraction ($\%$) \\
 \hline
 GMRT      			& 2800    & 100$\%$      \\
 SERVS band 1 or 2        	& 2369    & 85$\%$      \\ % Separation > 0
 SERVS band 1 and 2        	& 2234    & 80$\%$      \\ % ID_1 > 0 && ID_2 > 0
 SWIRE all IRAC bands    		& 1091    & 39$\%  $    \\ % 4 Fluxes > 0
 MIPS 24 $\mu$m			& 1672    & 60$\% $    \\ % FLUX_APER_? > 0 
 X-RAY        & 92      & 3.3$\%$ \\
 MRR13-PHOTZ        & 1456    & 52$\%$ \\ % condition from stilts command!
 MRR13-FIR-LUM         & 1279    & 46$\%$ \\
 SPECZ              & 817     & 29$\%$ \\
 REDSHIFTS              & 1760    & 63$\%$ \\ 
 \hline
 \end{tabular}
 \label{tab_matches} 
 \end{table}
 
\section{Redshifts}\label{redshifts.sec}

The majority of the spectroscopic redshifts for our sample were obtained with BOSS. The BOSS spectral 
classification and redshift analyses are based on a $\rm{\chi ^{2}}$ minimization of 
linear fits to each observed spectrum using multiple templates \citep{2012AJ....144..144B}. 
The BOSS spectroscopic redshifts were supplemented by a small number of redshifts available from the literature. 
For sources where a spectroscopic redshift was not available, we use photometric redshift estimates
from the revised SWIRE Photometric Redshift Catalogue \citep{RowanRobinson2013}. The SWIRE catalogue takes 
into account new optical photometry in several of the SWIRE areas, and incorporates Two Micron All Sky Survey 
(2MASS) and UKIRT Infrared Deep Sky Survey (UKIDSS) near-infrared data. It is the most reliable photometric 
redshift catalogue publicly available in ELAIS N1, and covers the full GMRT mosaic.

\begin{figure}
  \includegraphics[width = 1.2\columnwidth]{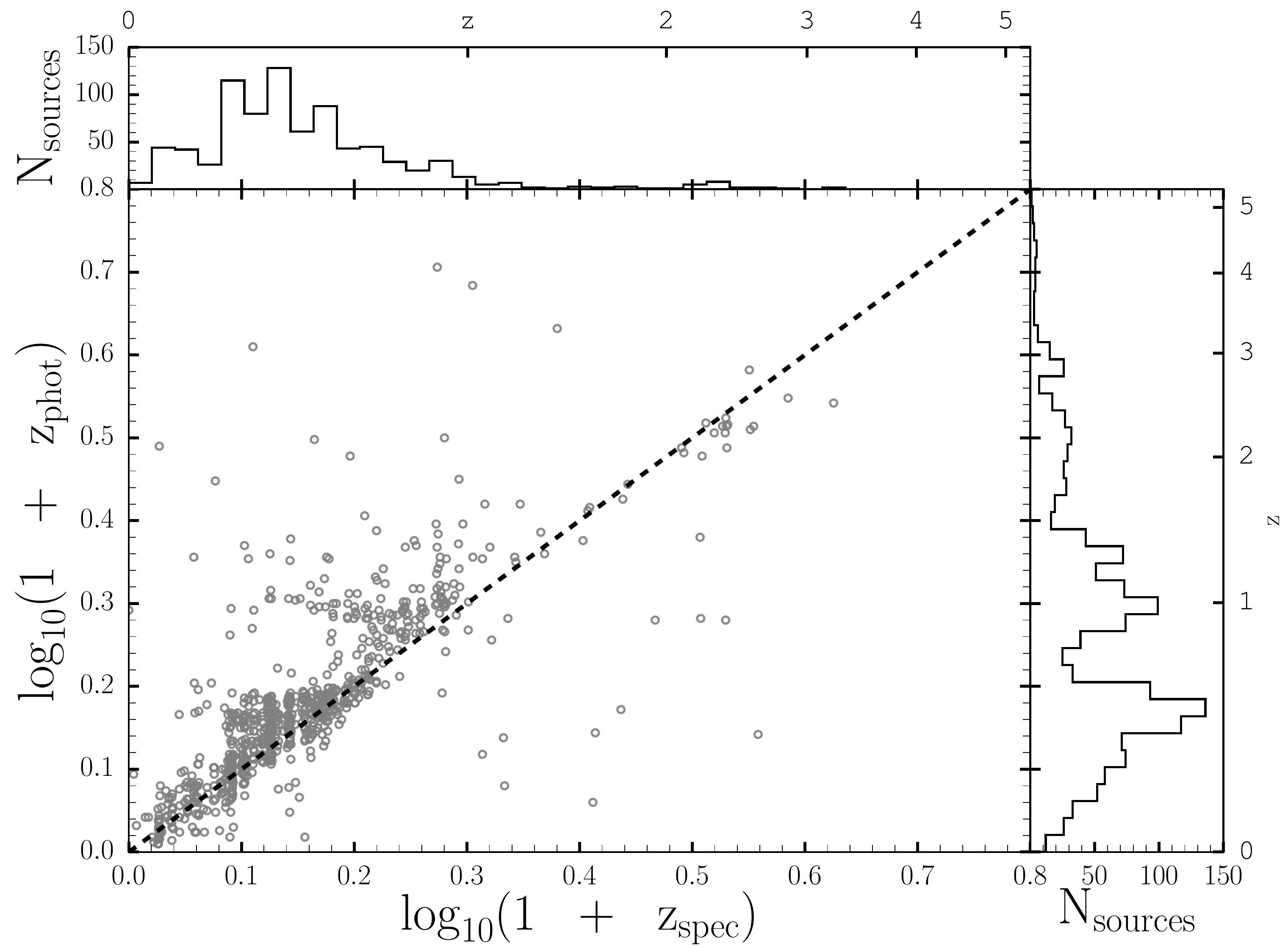}
  \caption{Comparison between photometric and spectroscopic redshift for sources with both. The dashed black line corresponds to $\rm{ z_{spec} \ = \ z_{phot}}$. The top and right histograms show spectroscopic and photometric redshifts respectively.} %The inset histogram show the distribution of the difference whereby the dash vertical lines indicate $|\Delta{z}|\,/(\,1\,+\,z_{spec})\,=\,0.2$. 
 % The mean $(\mu)$, and standard deviation $(\sigma)$ of the distribution are indicated in the top right of the panel}
  \label{fig_speczvsphotz} 
\end{figure}

The redshift statistics for our sample are summarized in Table~\ref{tab_matches}.
Spectroscopic redshifts are available for 29$\%$.
The SWIRE catalogue provides photometric redshifts for 52$\%$ of our sample and far-infrared luminosities 
for 46$\%$. 
Combined 63\% of our sample has either a spectroscopic or photometric redshift.
Figure~\ref{fig_speczvsphotz} compares photometric and spectroscopic redshifts for sources with
both. The histograms at the top and on the right-hand side of the figure show the distributions
of photometric and spectroscopic redshifts respectively. In general, the two redshifts agree out to spectroscopic 
redshift $z_{\rm spec} < 1$ (i.e. $\log_{10}(1 + z) < 0.3$). However, a non-negligible number of outliers can be 
seen creating a spurious peak in the photometric redshift distribution at $z_{phot} \sim 1$.

\begin{figure}
  \includegraphics[width =\columnwidth]{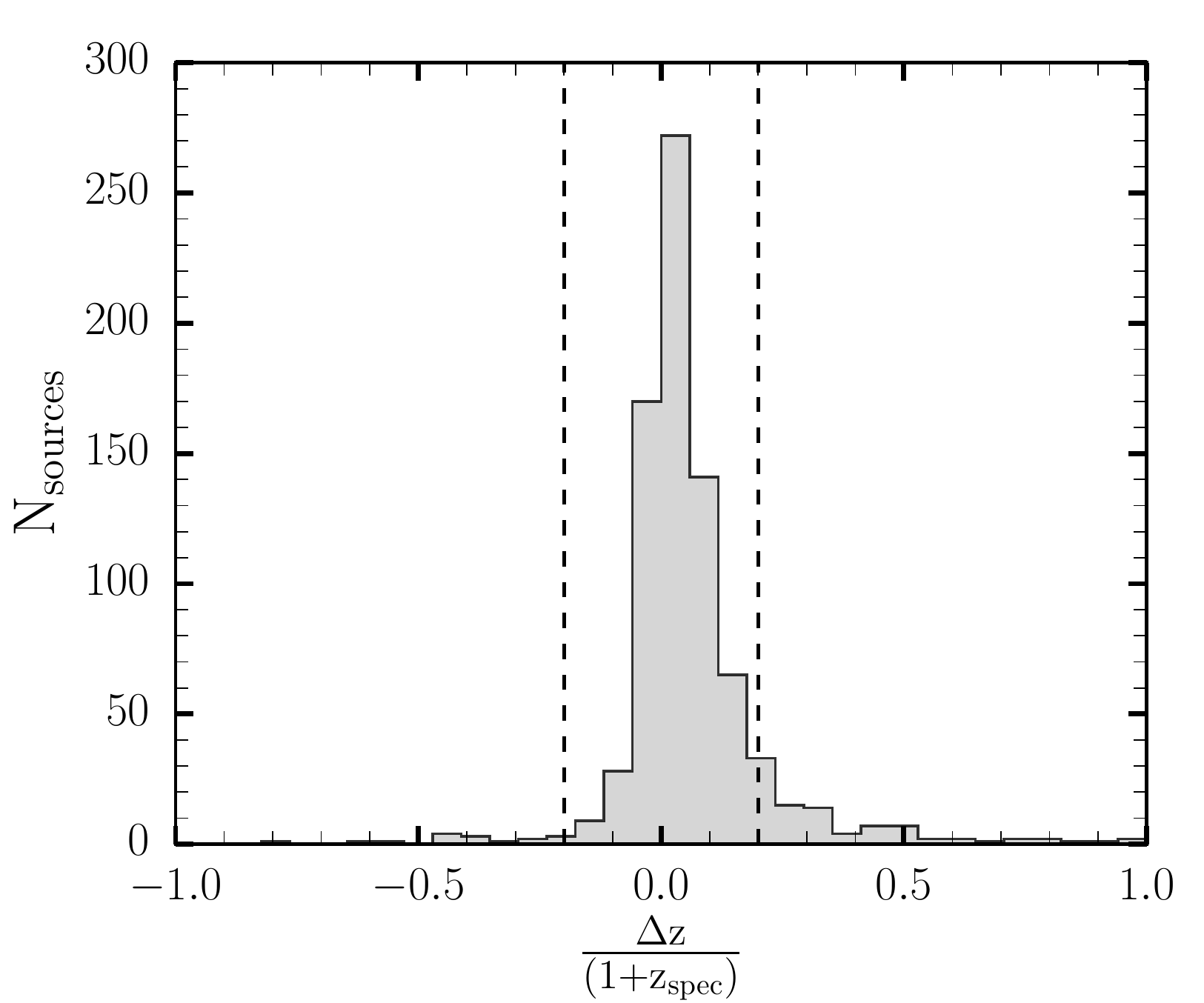}
  \caption{Histogram showing the distribution of the $|\Delta{z}|\,/(\,1\,+\,z_{spec})$ the dash vertical lines indicate $|\Delta{z}|\,/(\,1\,+\,z_{spec})\,=\,0.2$. 
  The mean $(\mu)$ and standard deviation $(\sigma)$ are 0.07 and 0.20 respectively.} 
  \label{fig_deltaz} 
\end{figure}

Figure~\ref{fig_deltaz} shows the distribution of the difference 
$(\Delta{z})/(1 + z_{\rm{spec}})$, where $\Delta{z} = z_{\rm phot} - z_{\rm spec}$.
The standard deviation of this distribution is $\sigma[\Delta z/(1 + z)] \sim 0.20$.
Outliers with $|\Delta{z}|\,/\,1\,+\,z_{spec}\,>\,0.2$ make up 10.40$\%$ of our sample.

We also compute the normalized median absolute deviation:
\begin{equation}\label{sigma}
\sigma_{\rm NMAD} = 1.48\times {\rm median} \left(\frac{|\Delta{z} - {\rm median}(\Delta{z})|}{1 + z_{\rm spec} }\right),
\end{equation}
an estimate of the quality of photometric redshift which is less sensitive to outliers \citep{2008ApJ...686.1503B}. 
We find $\sigma_{\rm NMAD} = 0.074$, which is slightly higher than found in other photometric redshift catalogues 
(e.g. see \citealt{2009ApJ...690.1236I,2009ApJS..183..295T,2010ApJS..189..270C}).

\section{AGN Diagnostics}\label{agn.sec}

Of our total sample of 2800 radio sources, 1760 have either a spectroscopic or a photometric redshift estimate. Out of these, 1526 have at least one multi-wavelength (i.e. not only based on radio properties) AGN diagnostic and constitute our final sample for the purposes of source classification. In this section we describe the combination of multi-wavelength AGN diagnostics employed to obtain a census of galaxies showing evidence of hosting an AGN within this sample.

\subsection{Radio Diagnostics}\label{radiosel.sec}

\begin{figure}
\includegraphics[width = \columnwidth]{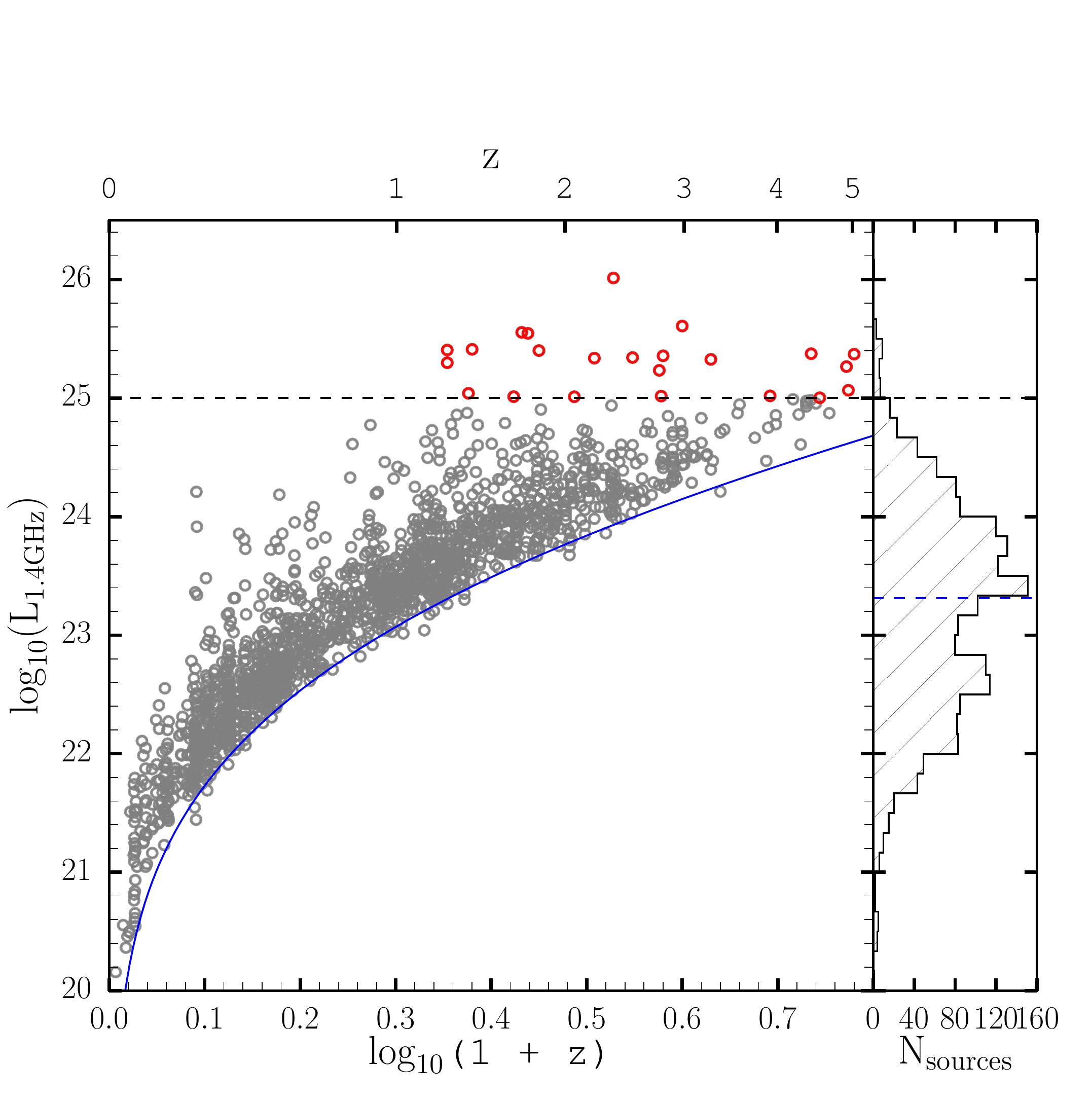}
 \caption{Radio Loud AGN Selection. The plot shows 1.4 GHz
  luminosity versus redshift for the GMRT sample with redshifts. The dashed black line shows a radio luminosity 
  threshold of $10^{25} \ {\rm W\,Hz^{-1}}$. Sources above this luminosity are classified as RL AGN (red circles). 
  The luminosity detection limit implied by the GMRT sensitivity is shown by the solid blue curve. The blue dashed line in the right hand histogram shows the median 1.4 GHz luminosity.}
\label{fig_radio} 
\end{figure}

The first criteria we use to identify AGNs in our radio sample are the radio luminosities. The 612 MHz radio flux densities were converted to rest-frame 1.4\,GHz effective luminosities assuming a radio spectral 
index of $\alpha = - 0.7$ \citep{2010MNRAS.401L..53I} as:
\begin{equation}\label{radiolum.eq}
L_{\rm 1.4\,GHz} \  = 4\pi d^{2}_{\rm lum} {\frac{S_{\rm 1.4\,GHz}} {(1 + z) ^{1 +\alpha}}}
\end{equation}
where
\begin{equation}
 {S_{\rm 1.4\,GHz}}\, = \,\left(\frac{1.4}{0.61}\right)^{\alpha}\,S_{\rm 0.6\,GHz}.
 \label{eq_1.4GHz}
 \end{equation}

We classify sources as RL AGNs based on a radio luminosity cut of $L_{\rm 1.4\,GHz} > 10^{25} \ {\rm W\,Hz^{-1}}$, as suggested by e.g., \cite{2007ApJ...667L..17S}; \cite{2007ApJ...656..680J} and \cite{2008ApJ...683..659S} and illustrated by Figure~\ref{fig_radio}. We note that this is a conservative criterion, as it only selects the brightest radio-loud galaxies and quasars. \cite{2007MNRAS.375..931M} showed that the AGN radio luminosity function flattens at low luminosities, indicating that there remains a fraction of AGN powered radio sources below the cutoff.  
 Our radio luminosity selection criterion, identifies 26 RL AGN sources, constituting 1.5\% of the population with redshifts.

\subsection{X-ray Diagnostics}\label{xraysel.sec}

\begin{figure}
\includegraphics[width = 0.49\textwidth]{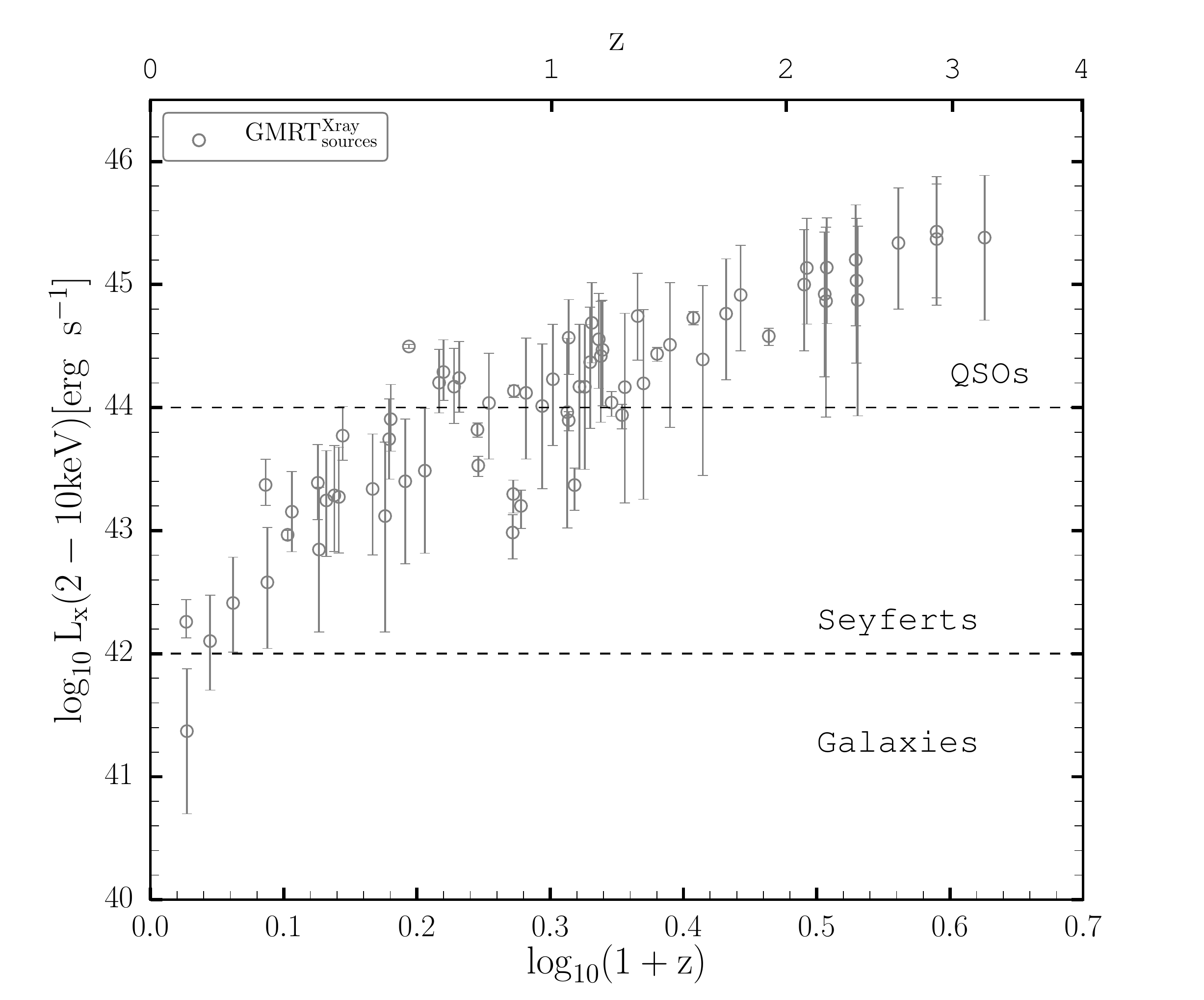}
\caption{X-ray luminosity as a function of redshift for the GMRT sources with X-ray detections. 
The dash horizontal line represents the typical demarcations for Galaxies, Seyferts and QSOs.}
\label{fig_xray} 
\end{figure}

X-ray surveys are perhaps one of the most reliable methods to select AGNs, since they directly probe their high energy emission. 
However dusty tori may obscure the X-ray emission and reprocess it into infrared emission. 
Thus, AGN detections through X-ray and infrared are largely complementary \citep{2014MNRAS.444L..95F}. 
We computed the hard X-ray luminosity (2 -10 keV) using the relation
\begin{equation}
L_{\rm x} \ = \ 4 \pi S_{\rm x} d_{\rm L}^{2} (1 +z)^{2 - \gamma}.
\end{equation}
where we fixed the  photon-index to the commonly observed value of $\rm{\gamma} = 1.8$ 
\citep{2008A&A...485..417D, 2014MNRAS.445.3557V}. 
Figure~\ref{fig_xray} shows the x-ray luminosity for the 70 objects having redshifts
that are detected in either the x-ray deep or x-ray shallow datasets.  
This includes only 1 of the 26 sources classified as RL AGN in the previous section.
We classify a source as an AGN based on x-ray emission 
when $L_{\rm x} > 10^{42}$ erg\,s$^{-1}$ following e.g. \cite{2004ApJS..155..271S}. 
All but one source of the 70 objects lies above this threshold.
We therefore classify 69 sources as hosting AGNs through this X-ray criterion,
representing 4.0\% of our radio sample with redshifts.  

\subsection{BOSS/SDSS Diagnostics}\label{bosssel.sec}
 
From our cross-matching, we were able to associate 779 of the GMRT sources with BOSS spectroscopic 
redshifts and classifications. The BOSS CLASS and SUBCLASS parameters, as detailed by 
\cite{2012AJ....144..144B},  classify spectra as:
\begin{enumerate}
\item[1.] GALAXY: identified with a galaxy template and can have the following subclasses:
\begin{itemize}
\item[a.] STARFORMING:  the spectrum has detectable emission lines that are consistent with star formation 
according to the criterion: \\
\indent $\log_{10}({\rm OIII/H}\beta)  <  0.7  -  1.2(\log_{10}({\rm NII/H}\alpha) \ + \  0.4) $
\item[b.] STARBURST: galaxy is star-forming with an equivalent width of H$\alpha$ greater than 50\,\AA
\item[c.]  AGN: the spectrum has detectable emission lines that are consistent with being a Seyfert or a LINER: \\
\indent $\log_{10}({\rm OIII/H}\beta)  >  0.7  -  1.2(\log_{10}({\rm NII/H}\alpha) \ + \  0.4) $
\end{itemize}
\item[2.] QSO: identified with a QSO template. \\
\item[3.] STAR: identified with a stellar template.  \\
\end{enumerate}
In addition, any galaxies or quasars that have lines detected at the 10$\sigma$ level with 
velocity dispersion $>$ 200 km\,s$^{-1}$ at the 5$\sigma$ level, have the category BROADLINE 
appended to their SUBCLASS.

The BOSS analysis of the 779 sources yields 703 GALAXY and 73 QSO classifications. The remaining 3 sources were identified as STARS. Among the 703 GALAXY CLASS, 19 are classified as SUBCLASS AGN.  Conversely, 172 sources are classified as STARFORMING and 141 classified as STARBURST. 
Four sources are classified are STARBURST BROADLINE and 43 sources classified as BROADLINE. 
The remaining 324 sources of the 703 GALAXY class had no SUBCLASS associations.

Of the three BOSS spectra identified as STARS their SUBCLASS parameter indicates that they are an M4.5:111, 
GOVa, and a cataclysmic variable star respectively.  We do not include these sources in our BOSS selection. 

We classify as AGN all objects with a BOSS QSO class, as well as any source having a BOSS GALAXY class with a 
SUBCLASS parameter of either AGN,  BROADLINE  or STARBURST BROADLlNE.  All others GALAXY class are taken to be SFG.
Table~\ref{tab_bossclass} presents the  breakdown of the number of the SFGs and AGNs from the BOSS spectroscopic classifications. 

\begin{table}
\caption{BOSS spectroscopic classifications results.}
\centering
\begin{tabular}{lrr}
\hline
\hline
Category          & Number & Fraction \\
\hline
SFG                  & 683 & 39.7$\%$ \\
AGN                  &  96 &  5.5$\%$ \\
\hline
\end{tabular}
\label{tab_bossclass} 
\end{table}
\subsection{IRAC Color Diagnostics}\label{iracsel.sec}
The IRAC four-band color-color plane was used to identify AGNs adopting the criterion proposed by 
\cite{2012ApJ...748..142D} and defined as:
\begin{equation}
{x \ = \ \log_{10}\left(\frac{f_{5.8\mu m}} {f_{3.6\mu m}}\right) , \    \ y = \log_{10}\left(\frac{f_{8.0\mu m}}{f_{4.5\mu m}}\right)}
\end{equation}
\begin{multline}
{x  \ge \ 0.08 \ \wedge \ y   \ge  0.15}
\\ 
{\wedge  y \ge  (1.21  \times   x)  -  0.27}
\\
{\wedge y \le  \ (1.21  \times  x)  + 0.27}
\\
{\wedge  f_{4.5\mu m}   >  f_{3.6\mu m}  >   f_{4.5\mu m}   \wedge   f_{8.0\mu m}   >   f_{5.8\mu m}} \\
\label{donley_crit}
\end{multline}

\begin{figure}
\includegraphics[width = 1.1\columnwidth]{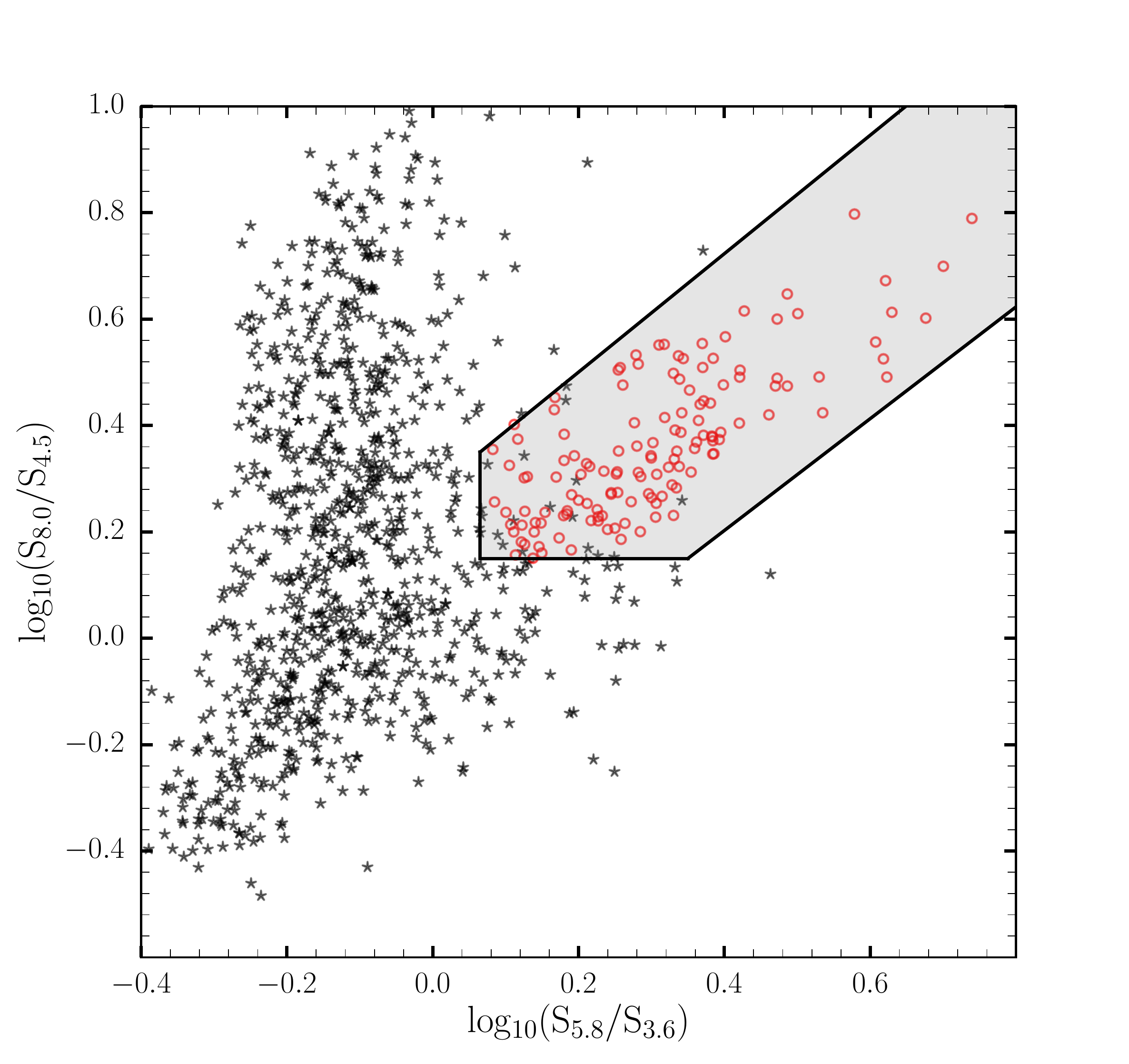}
\caption{IRAC color-color diagram showing the separation between AGNs (open red circles) and SFGs (black stars). 
The shaded region indicates the area within the Donely wedge.}
\label{fig_irac} 
\end{figure}

This criterion have been designed to reject the majority of low and high redshift star-forming contaminants  
in the \cite{2004ApJS..154..166L, 2007ApJ...669L..61L} and \cite{2005ApJ...631..163S} AGN selection wedges. 
Figure~\ref{fig_irac} presents IRAC color-color diagram for our 1091 GMRT sources with four IRAC band detections. We select 138 sources from this criterion. 
Of these, 123 sources have redshift estimates, constituting $7.0\%$ of the radio sources with redshift. 

\subsection{Mid-Infrared - Radio Flux Ratio}\label{mipssel.sec}

The MIPS \citep{2004ApJS..154...25R} 24$\mu{\rm m}$ flux density and the effective 1.4 GHz flux density, 
$S_{\rm 1.4\,GHz}$ (Equation \ref{eq_1.4GHz}), were used to calculate the $q_{24\mu{\rm m}}$ parameter,
\begin{equation}\label{qobs}
q_{24\mu{\rm m}} = \log_{10}(S_{24\mu {\rm m}} / S_{\rm 1.4\,GHz}).
\end{equation}

%Within our sample, 50\% of GMRT sources with MIPS 24$\mu{\rm m}$ detections have redshift estimates.
We compute $q_{24\mu \rm m}$ for the radio sources with MIPS 24$\rm{{\mu m}}$ detections and redshifts and compare it to the redshifted $q_{24\mu {\rm m}}$ value for the M82 local standard starburst galaxy template as done by \cite{2013MNRAS.436.3759B}  to identify SFGs and radio-loud AGNs. 
We normalize the M82 SED to the local average value of 
$q_{24\mu {\rm m}}$ obtained by \cite{2010ApJ...714L.190S} and define the SFG locus as the region within 
$\pm2\sigma$ of the normalized M82 template in the $(q_{24\mu {\rm m}},z)$ plane, where $\sigma$ is the 0.35 
average spread for local sources by \cite{2010ApJ...714L.190S}.
Sources below this locus display a radio excess with respect to SFGs and are therefore classified as 
Radio-Loud AGNs. 

\begin{figure}
\includegraphics[width = 1.05\columnwidth]{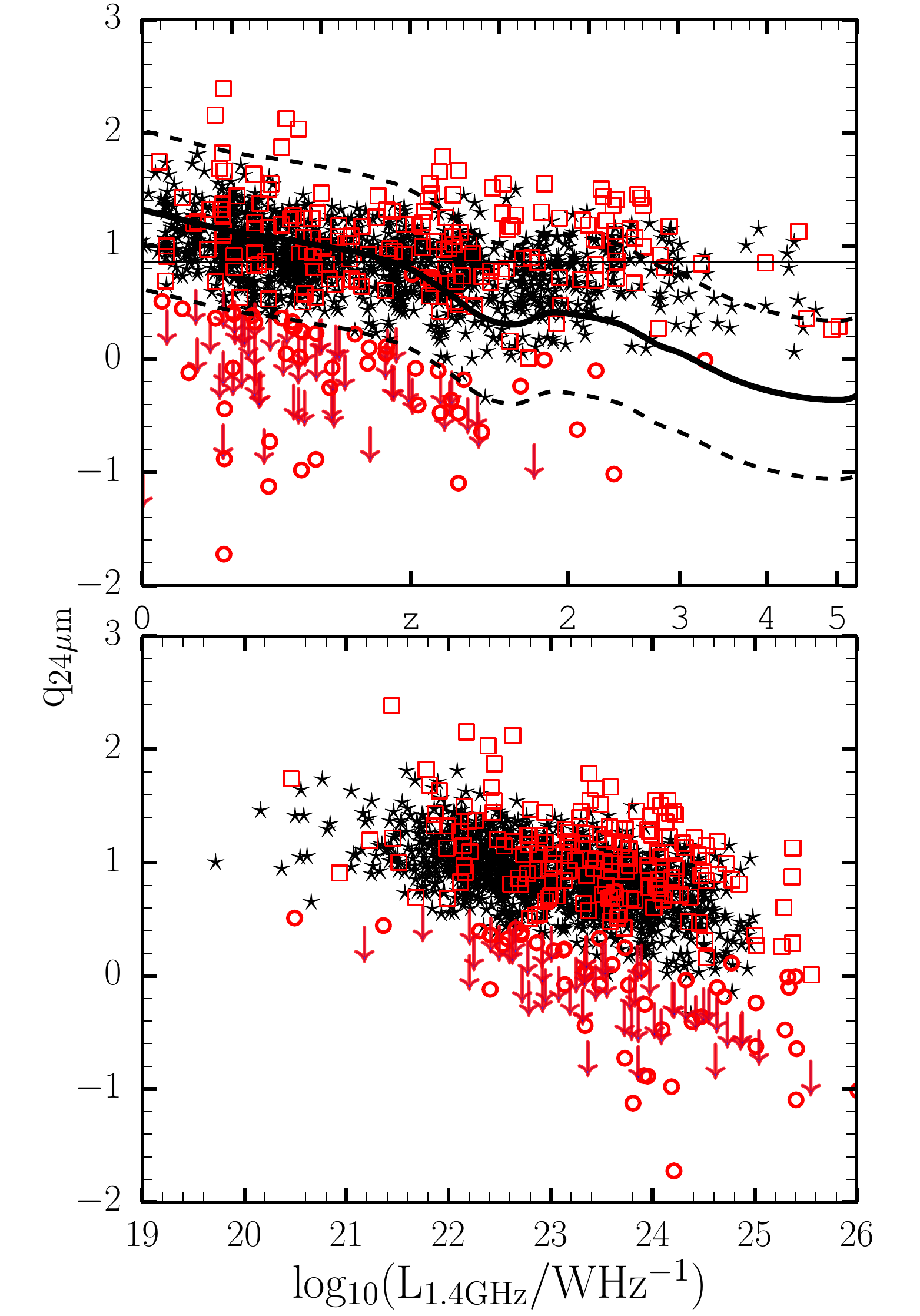}
\caption{Top: $q_{24\mu {\rm m}}$, the logarithm of the ratio between the mid-infrared and radio flux for the GMRT 
sources with redshift estimates. The solid black curve shows the predicted value of M82 template, the dividing line between 
radio-quiet and radio-loud AGNs with $\pm2\sigma$ dispersion (dash black curve). The horizontal line is the median value 
of $q_{24\mu {\rm m}}$ for the sources. The black stars, open red squares and open red circles  SFGs, RQ AGN and RL AGN respectively. The downward pointing arrows represent RL AGN with upper limits (see Equation~\ref{qobs_limit})).  Bottom: $24\mu {\rm m}$ vs. 1.4\,GHz luminosity for SFGs, RQ AGNs and  RL AGNs.}
\label{fig_qobs} 
\end{figure}

The results for our sources are illustrated in the top panel of Figure~\ref{fig_qobs}. The black stars, open red squares above the $-2\,\sigma$ dispersion curve (dash black curve) represents the SFGs and RQ AGN respectively.
This criterion, initially selects 42 radio sources as RL AGN (open red circles in Figure~\ref{fig_qobs}) constituting $\sim\,2.4\%$ of the radio population with redshifts.
We compare this method to the radio-loud AGN classification based on radio luminosity in Section~\ref{radiosel.sec}.
Of the 26 sources selected as radio-loud AGNs according to their radio luminosity 15 have mid-infrared detections. Amongst these, only 5 are also classified as radio-loud using the $q_{24\mu \rm m}$ selection criterion.
For the sources with a redshift detection but no MIPS counterpart, we assumed that:

\begin{equation}\label{limit}
S_{24\mu {\rm m}} < S_{24\mu {\rm m}\,{\rm lim}} = 286.6 \mu\,Jy
\end{equation}

and thus the upper limit (downward pointing arrow symbols in Figure~\ref{fig_qobs} is given by:
\begin{equation}\label{qobs_limit}
q_{24\mu{\rm m}\,{\rm lim}} = \log_{10}(S_{24\mu {\rm m}\,{\rm lim}} / S_{\rm 1.4\,GHz})
\end{equation}

For our sources with redshifts within the same range studied by \cite{2010ApJ...714L.190S}, \textbf{i.e $z \sim 1.4$}, we measure a 
median $q_{24\mu {\rm m}}$ value of $1.10 \pm 0.02$, whereas for our full sample with redshifts we measure 
a median $q_{24\mu \rm m}  =  0.86 \pm 0.01$ at a median redshift of $\sim 0.71$. 
We estimated the median 
and the error on the median using the median absolute deviation estimator as this is 
a more robust measure of the variability of a univariate sample of quantitative data than the standard deviation
 \citep{doi:10.1080/01621459.1993.10476408}.  Our result is in agreement with and more precise than 
 previous work from \cite{2004ApJS..154..147A} 
 who estimated $q_{24\mu \rm m} = 0.84 \pm 0.28$ by matching over 500 Spitzer sources at 24$\mu{\rm m}$  with VLA 1.4 GHz $\rm{\mu Jy}$ radio sources for the Spitzer First Look Survey (FLS) extending to $z > 2$. 
 \cite{2010ApJ...723.1110H} measured a median $q_{24\mu \rm m}  =  0.71 \pm 0.31$ for 84 sources with detections in both the IR and radio for observations of the Hubble Deep Field South taken with the Spitzer Space Telescope up to $z > 1$.
\subsection{AGN/SFG Classification Overview}\label{diagnostics.sec}
\begin{figure}
\includegraphics[width =1.12\columnwidth]{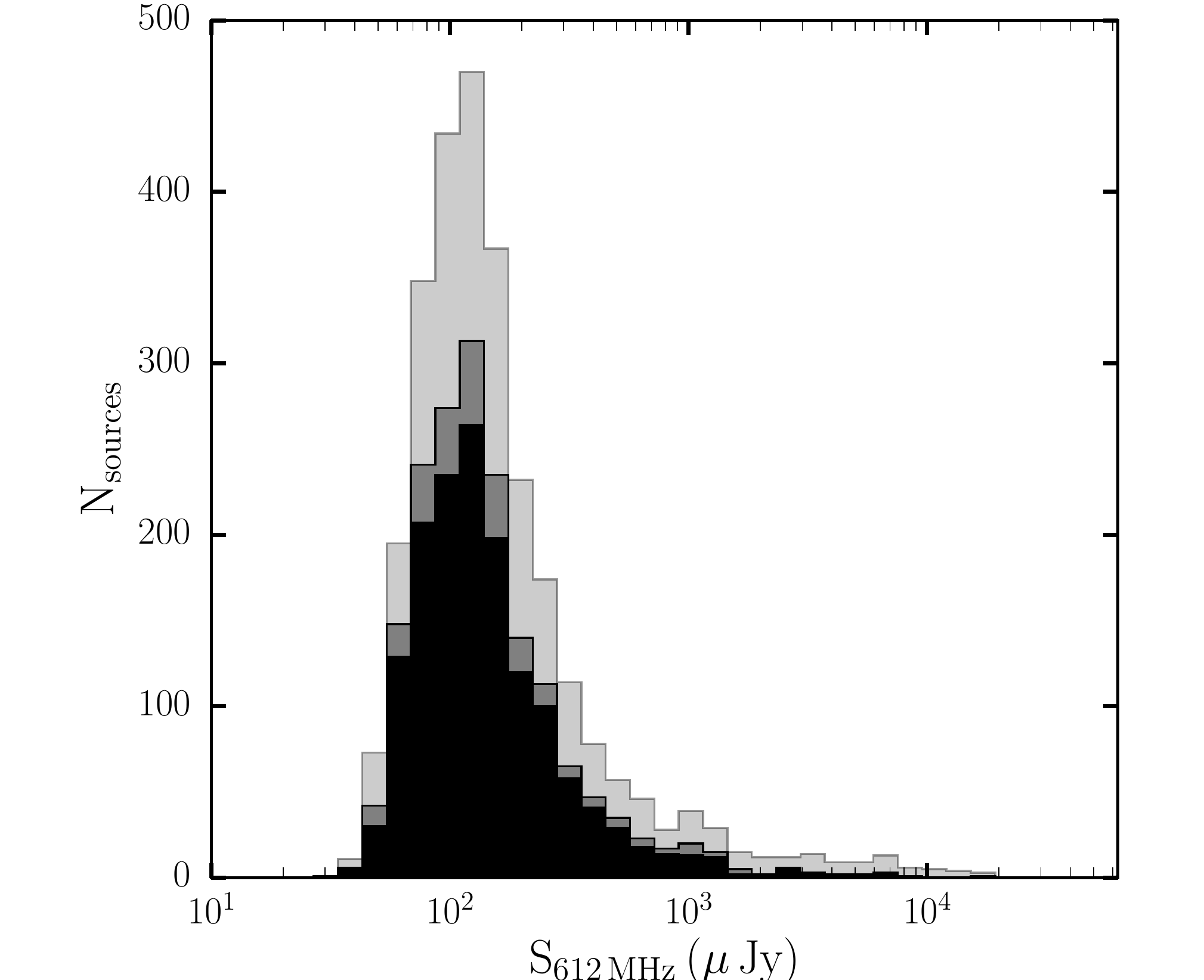}
\caption{The distribution of 612-MHz flux densities for the entire GMRT sample of 2800 sources (light gray), radio sources with redshifts (dark gray) and sources with redshifts that also have at least one diagnostic for AGN activity (black).}
\label{fluxdist.fig}
\end{figure}

\begin{figure}
\includegraphics[width =1.15\columnwidth]{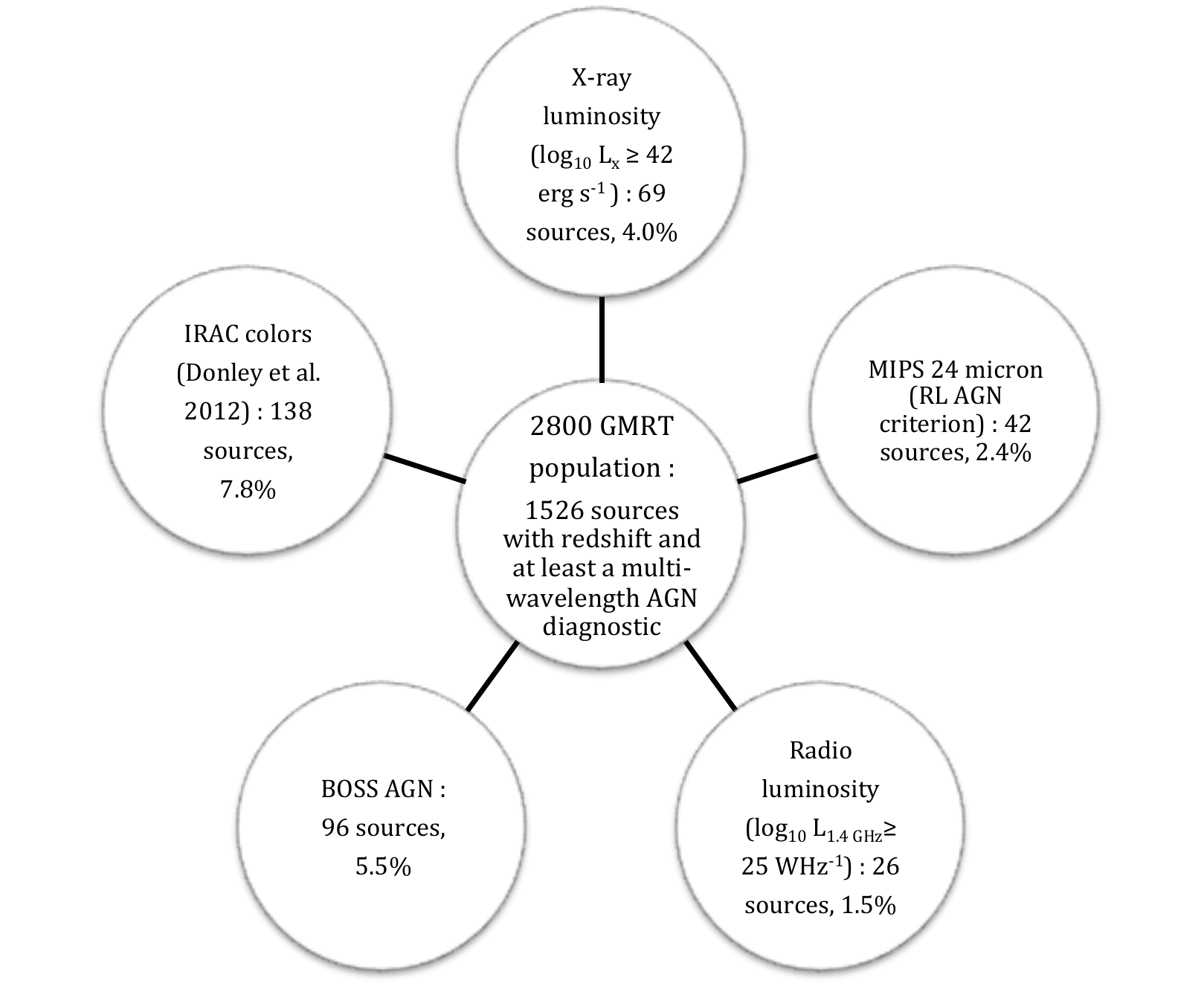}
\caption{A schematic of the various AGN selection.}
\label{diagnostics.fig}
\end{figure}

We have carried out a multi-wavelength study using optical, X-ray, infrared and radio data to search for evidence of the presence of an AGN in our sample of 1526 radio sources using five criteria:

\begin{enumerate}
	\item Radio power, $L_{\rm 1.4\,GHz} \ge 10^{25}$  W\,Hz$^{-1}$.
	\item X-ray luminosity, $L_{\rm x} \ge 10^{42}$ erg\,s$^{-1}$.
	\item BOSS AGN spectroscopic classification.
	\item IRAC colors using the \cite{2012ApJ...748..142D} criteria.
    \item Mid-Infrared to radio flux ratio using the \cite{2013MNRAS.436.3759B} criterion.
\end{enumerate}
\noindent

%Of the 1760 objects with redshifts we are able to apply at least one multi-wavelength diagnostic to \textbf{1526} radio sources.
The 612-MHz flux density distribution of the sample is shown in Figure~\ref{fluxdist.fig}. The entire range of flux densities in the total GMRT population is sampled.  The minimum flux density for the 1526 sources with AGN diagnostics is 52\,$\mu$Jy and the median flux density is 123.3\,$\mu$Jy, close to the median of 128.8\,$\mu$Jy for
the total sample.

Combining all the indicators we identify sources that exhibit evidence of the presence of an AGN. The breakdown of classification from each indicator is illustrated in Figure~\ref{diagnostics.fig}, and Table~\ref{indicators.tab} provides a matrix showing the correlation of classification among the different indicators.
The most successful indicators, in terms of the number of objects identified as AGN, are the IRAC colours and BOSS spectral classification.
We note that only one source from the Radio power AGN criterion out of a total of 26 sources, has an X-ray luminosity greater than $\rm{10^{42} \ erg \ s^{-1}}$. 
Also, two of these sources are confirmed as AGNs by the BOSS criterion whereas 6 of them are identified by the IRAC criterion. 
The disagreement between the radio luminosity and the X-ray luminosity criterion may be attributed to heavily obscured sources in the X-rays, since radio observations are almost unaffected by dust extinction and therefore can detect even the most obscured systems \citep{2013MmSAI..84..665D}.   

\begin{figure}
\includegraphics[width =1.0\columnwidth]{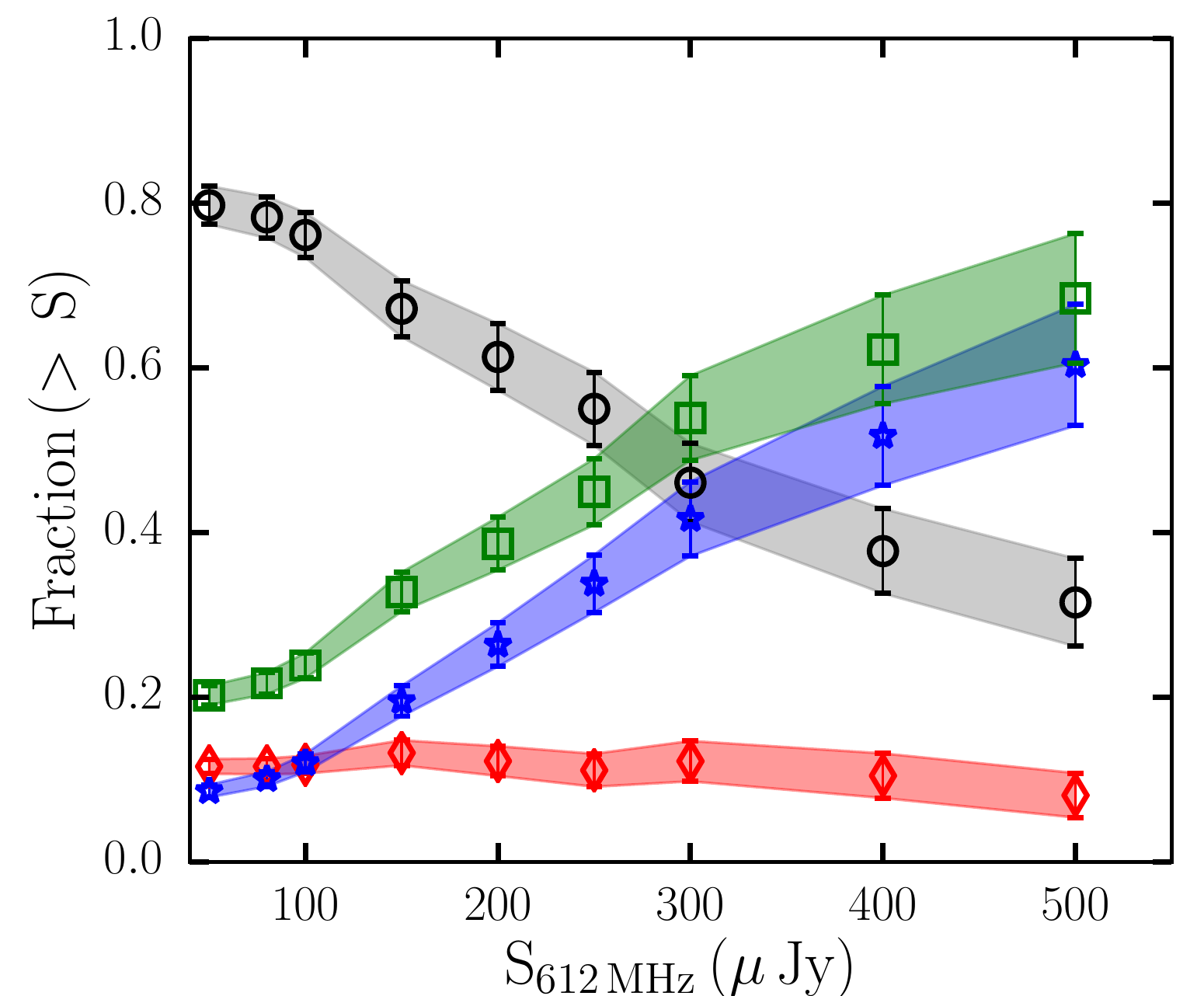}
\caption{The relative fraction of AGN and SFG as a function of minimum flux density.
The data points show the fraction of the sample greater than flux density, $S_{\rm 612\,MHz}$, that are classified as RQ\,AGN (red diamonds), RL\,AGN (blue stars), all AGN (green squares) and SFG (black circles).
}
\label{fractions.fig}
\end{figure}
  
\begin{table*}
\caption{Number of sources identified as AGN by each of the diagnostics. Entries indicate the number of AGN that are common to both indicators.}
\centering
\begin{tabular}{lccccc}
\hline
\hline
Indicator & Radio Luminosity  & X-ray   & BOSS AGN  & IRAC colors  & MIPS 24  \\
\hline
Radio luminosity  &  26        &  -          &  -              &  -                & -           \\
X-ray        &   1             & 69          &  -              &  -                & -           \\
BOSS AGN	 &   2             & 26          & 96              &  -	               & -           \\
IRAC colors	 &   6             & 35          & 45              & 138	           & -           \\
MIPS 24	     &   5             &  5          &  5              &   2               & 42          \\
\hline
\end{tabular}
\label{indicators.tab}
\end{table*}

We further classify the AGNs as RL AGN for all the sources with either $L_{\rm 1.4\,GHz} > 10^{25} \ {\rm W\,Hz^{-1}}$ (see Section~\ref{radiosel.sec}) or $q_{24\mu {\rm m}}$ below the M82 locus (see Section~\ref{mipssel.sec}). Above this threshold, a source is classified as a RQ AGN if it shows clear evidence of an AGN in the X-ray, in the BOSS/SDSS AGN spectroscopic classification, from its radio power or from its IRAC colors satisfying the Donley criterion. Table~\ref{classification.tab} presents the total number of AGNs and SFGs from a combination of the selection criteria for sources with redshifts.
The objects classified as SFG are those sources in our redshift sample that do not show evidence of AGN activity in any of the diagnostics.   
 This may be considered an upper limit to the population of sources in this flux density regime whose radio emission is powered by star formation processes. The large number of objects in our sample allows us to examine the change in population with flux density.
 Figure~\ref{fractions.fig} shows the fraction of objects in each class in our sample as a function of limiting flux density.  For a given flux density, $S_{\rm 612\,MHz}$, the plot shows the fraction of objects that are classified as SFG, RL\,AGN, and RQ\,AGN in the sample of objects above that flux density.  
 The green points shows the fraction for the total AGN population. 
 The curves highlight the dramatic change in population over this flux density range.
 The SFG fraction exhibits a monotonic increase with decreasing flux density from $\sim$30\% to 80\%.   
 RL\,AGN decrease rapidly from being the dominant population above $\sim$400\,$\mu$Jy to the smallest fraction below $\sim$100\,$\mu$Jy.
 The fraction of RQ\,AGN remains roughly constant with flux density at $\sim$10\%.
 
\begin{table}
\caption{Total number of SFGs, RQ AGNs and RL AGNs from a combination of the selection criteria for sources with redshift.}
\centering
\begin{tabular}{llr}
\hline
\hline
Class & Number  &Fraction  ($\%$)\\
\hline
SFG    & 1226      & 80.3    $\%$        \\
RQ AGN & 173       & 11.4    $\%$        \\
RL AGN & 127       & 8.3     $\%$        \\
\hline
\end{tabular}
\label{classification.tab} 
\end{table} 

\cite{2013MNRAS.436.3759B} found that $\sim60\%$ of the 883 radio sources above $\rm{\sim 50 \mu Jy}$ 
from a VLA image at 1.4 GHz on $\sim0.3$\,deg${^2}$ of the ECDFs to be SFGs, 
$\sim26\%$ to be RQ AGNs and $\sim14\%$ to be RL AGNs.
\cite{2015MNRAS.452.4111R} separated the same 1.4 GHz extragalactic radio source population as \cite{2013MNRAS.436.3759B} into AGN and SFGs by fitting IR SEDs constructed using IRAC, MIPS and SPIRE photometry. Overall, they found that for 40\% of their sample the radio emission is powered by star formation, while for 20\% of their sample the radio emission is powered by an AGN and the remaining 40\% of their sample is made up by hybrids (i.e. no sign of jet activity, not clear whether or not the radio emission is powered by AGN).

Using a smaller but deeper 1.4 GHz sample reaching a 32.5\,$\mu$Jy flux limit over 0.29\,deg${^2}$ of the ECDFS VLA image, \cite{2015MNRAS.452.1263P}  
identified 626 radio sources with redshifts and classified 55\%, 25\%  and 20\% as SFGs,  RQ AGNs and RL AGNs respectively. \cite{2016MNRAS.462.2934V} classified 10\% RL AGN, 28\% RQ AGN and 68\% SFGs from 558 sources at 3\,GHz in the Lockman Hole North 
from a single pointing with the JVLA. 
 
For a spectral index of $-0.7$, the equivalent 1.4\,GHz flux density threshold of our radio selected sample  is 29\,$\mu$Jy, 
significantly deeper than  the Bonzini et.\ al.\ study, and slightly deeper than Padovani et.\ al. The increased depth combined with our larger survey area provides a source sample that is 70\% larger than the Bonzini study and over twice the size of the Padovani sample. 
In comparison to Bonzini et.\ al.\, our results indicate that the fraction of radio sources attributed to SFGs continues to increase rapidly with decreasing flux density, from $\sim$60\% to 80\%.   
We also confirm that RQ AGN dominate over RL AGN, with 40\% more RQ objects 
compared to RL objects in our sample. 

We find a much higher fraction of SFG compared to the analysis of Padovani et.\ al., 
which found instead approximately equal fractions of SFG and AGN.
This difference may be partially explained by the lower frequency of the GMRT sample, which selects against faint inverted spectrum sources, such as associated with compact AGN cores, that may be present in the VLA sample near the 1.4\,GHz flux threshold.  
Future studies of the spectral index distribution of $\mu$Jy radio sources will be important to resolve this difference.

\section{Discussion}\label{results.sec}

\subsection{Properties of the classified sources}

\begin{figure}
  \includegraphics[width = 0.9\columnwidth]{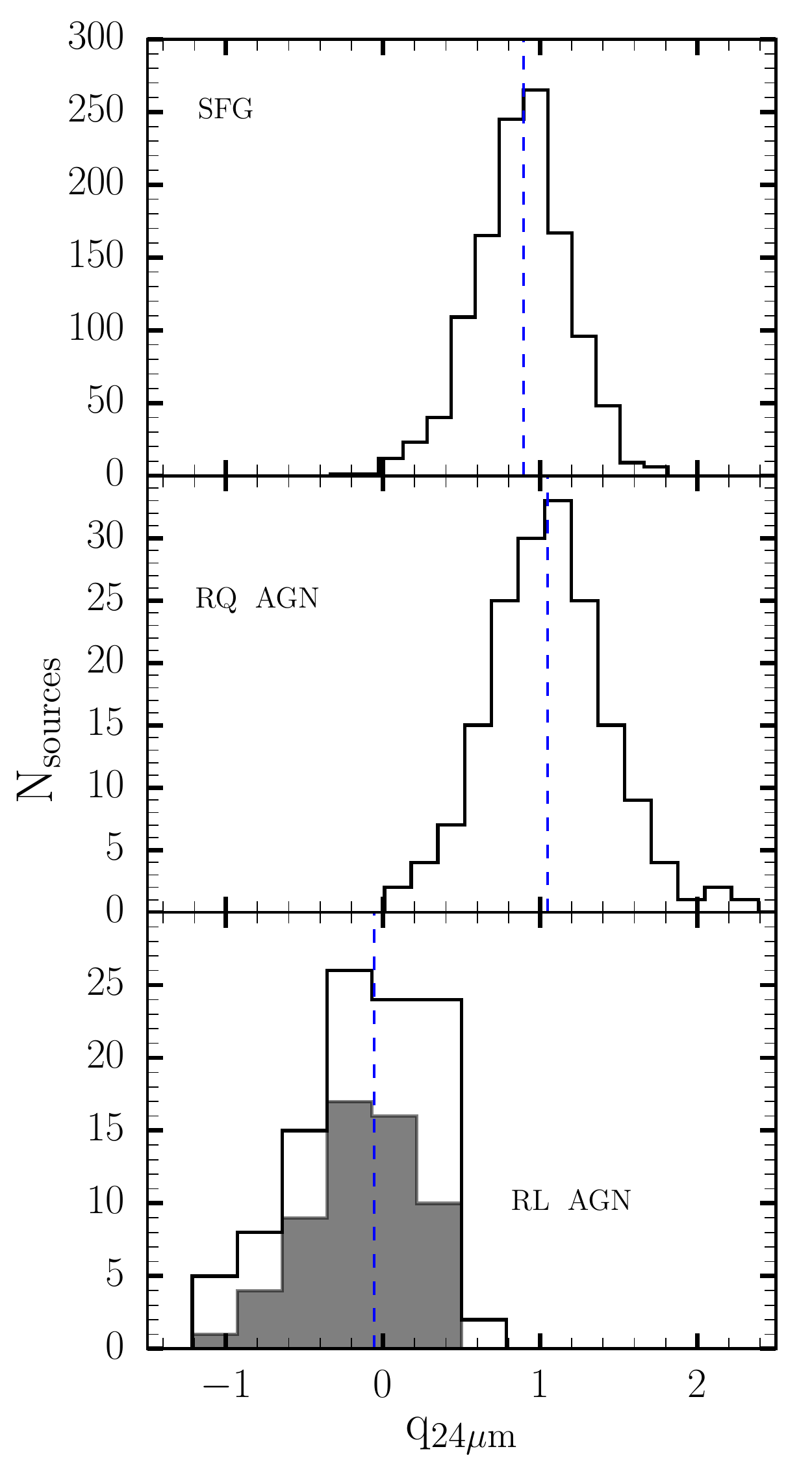}
  \caption{Distribution of the $q_{24\mu \rm m}$-parameter for our selected SFG (top panel), RQ AGN (middle panel) and RL AGN (bottom panel). The shaded histogram in the bottom panel show the $q_{24\mu \rm m}$ distribution for the RL AGNs selected from the upper limits given in equation~\ref{qobs_limit}  The dash vertical line in each panel designates the median value of the entire distribution, also listed in each panel.}
   \label{gm_q24_sfg_rq_rl.fig}
\end{figure}

\begin{table*}
\caption{Median values of redshift ($z$), $L_{\rm 1.4\,GHz}$, $L_{\rm IR}/L_{\odot}$ and $<q_{24\mu \rm m}>$ for the different class of sources.}
\centering
\begin{tabular}{lcccr}
\hline
\hline
Class  &$<z>$  &$<\log_{10}(L_{\rm 1.4\,GHz})>$ &$<\log_{10}( L_{\rm IR}/L_{\odot})>$ &$<q_{24\mu \rm m}>$ \\
\hline
SFG     &0.81 $\pm$ 0.01     &23.24 $\pm$ 0.02    &12.44 $\pm$ 0.04 & 0.89$\pm$ 0.01\\
RQ AGN   &1.09 $\pm$ 0.07     &23.54 $\pm$ 0.07   &12.61 $\pm$ 0.09 & 1.05$\pm$ 0.03\\
RL AGN  &0.88 $\pm$ 0.08    &23.92 $\pm$ 0.13    &13.02 $\pm$ 0.39 & -0.06$\pm$ 0.07\\
\hline
\end{tabular}
\label{tab:sfg_rq_rl_med} 
\end{table*}

The bottom panel of Figure~\ref{fig_qobs} plots $q_{\rm 24\mu m}$ as a function of 1.4 GHz luminosity for SFGs (black stars), RQ AGNs (open red squares) and RL AGNs (open red circles). The $q_{\rm 24\mu m}$ parameter seems to show a decreasing trend with increasing radio luminosity, and this is clearly evident for RL AGNs.
In Figure~\ref{gm_q24_sfg_rq_rl.fig} we show the distribution of $q_{\rm 24\mu \rm m}$for  
SFGs (top panel),  RQ AGNs (middle panel) and RL AGNs (bottom panel). 
The median $q_{24\mu \rm m}$ value for SFGs is  $0.89 \ \pm \ 0.01$ with a distribution that is peaked at the median with an RMS scatter of 0.18. 
The median $q_{24\mu \rm m}$ for RQ AGN is larger than for SFG at $1.05 \ \pm \ 0.03$
but like SFG is also peaked at the median with a slightly larger RMS of 0.24. 
RL AGN, on the other hand, have much lower median $q_{24\mu \rm m}$ of  $-0.06 \ \pm \ 0.07$, and a broader, less peaked distribution with a larger RMS scatter of $0.34$.

\begin{figure*}
\includegraphics[width = 0.9\textwidth]{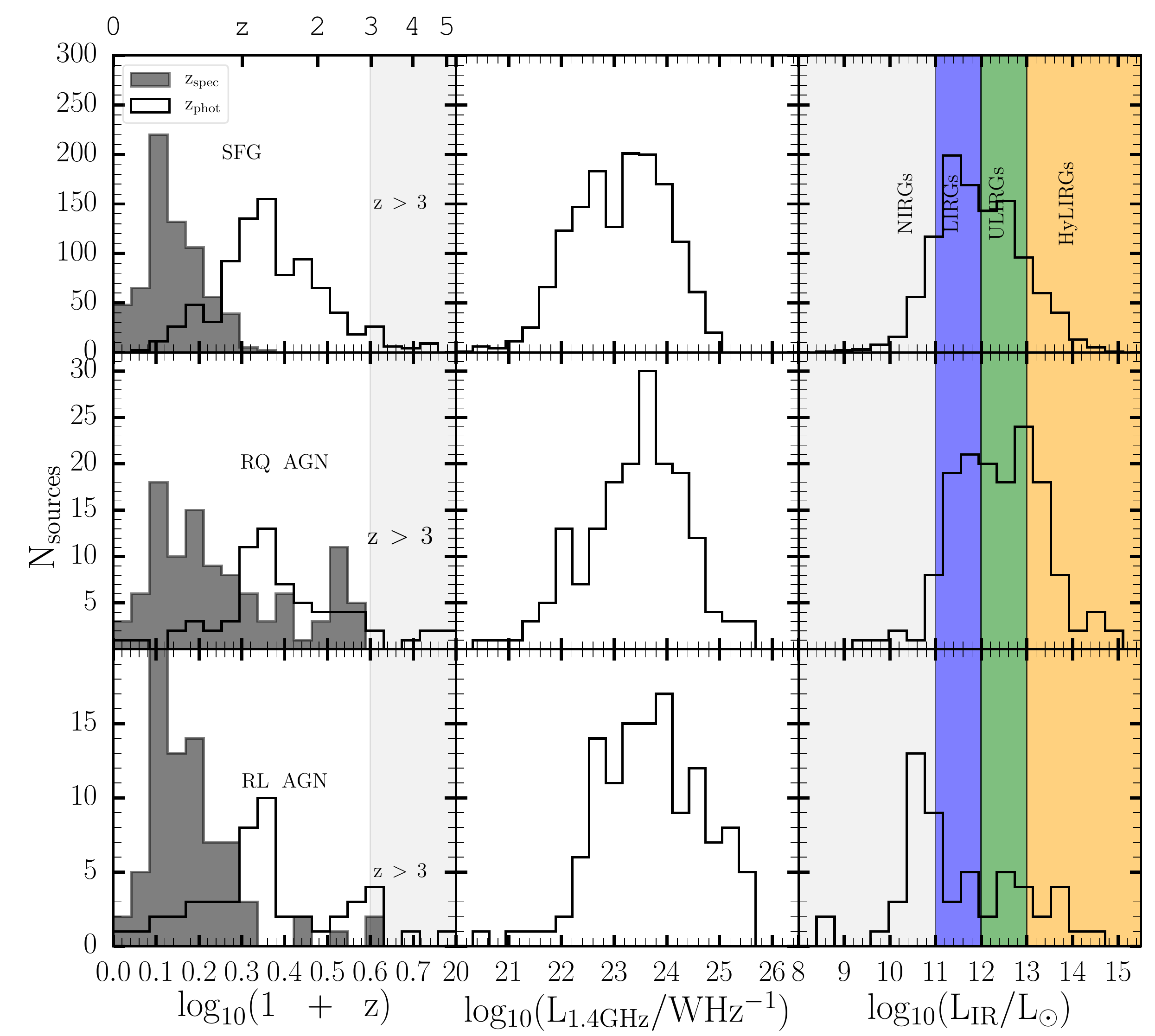}
\caption{Distribution of SFG (top panels), RQ AGN (middle panels) and RL AGN (bottom panels) for their redshifts, 
$\rm{L_{1.4 \ GHz}}$ and FIR luminosities respectively. The left vertical panels represent the distributions for 
spectrocopic (filled) and photometric (empty) redshifts for the various class of sources. The grey shaded areas show $z>3$. 
The middle vertical panels represent the radio power distributions.
The limits of NIRGs, LIRGs, ULIRGs and HyLIRGs  \citep{2013MNRAS.428..291P} are shown as grey, blue, green and orange shaded areas respectively in the right vertical panels to represent the range of sources based on their 
FIR luminosity. } 
\label{gm_sfg_rq_rl_hist.fig} 
\end{figure*}

Far-Infrared luminosities, $L_{\rm IR}$,  were derived from the integrated (8 - 1000$\rm{\mu m}$, rest frame)  far-infrared luminosities estimated by \cite{RowanRobinson2013} (RR13) using SWIRE photometry.
We corrected the RR13 values to our more accurate spectroscopic redshift by 
 \begin{equation}\label{corr}
L_{\rm IR}  =  \frac {d_{\rm L}^{2}}{ d_{\rm L, RR13}^{2} } L_{\rm IR, RR13}.
\end{equation}
Here  d$\rm{_{L,}^{2}}$ is the luminosity distance calculated using our redshift and 
d$\rm{_{L, RR13}^{2}}$ is the luminosity distance using the RR13 redshifts.	

The distributions of redshift, $L_{\rm 1.4\,GHz}$ and FIR luminosities for the three populations  are presented in 
Figure~\ref{gm_sfg_rq_rl_hist.fig}. 
The redshift distributions are shown in the left vertical panels. 
Photometric (empty) and spectroscopic (filled) redshift distributions are shown separately and grey shaded area 
indicates $z > 3$. 
We detect objects of all three types over the full redshift range up to $z > 3$.
The median redshift for each is $\sim$1.  Precise values are listed in Table~\ref{tab:sfg_rq_rl_med}.

The middle vertical panels of Figure~\ref{gm_sfg_rq_rl_hist.fig} show the radio power distribution for the three class of sources . 
As for $q_{\rm 24\mu m}$ SFG and RQ AGN show peaked distributions while the RL AGN have a wider distribution without a clear maximum.
The  distributions also show a consistent increase in the median radio power from SFG to RQ AGN to RL AGN.
As shown in Table \ref{tab:sfg_rq_rl_med} the increase in the median is  
$\Delta( \log{L_{\rm 1.4 \rm GHz}}) = 0.30 \pm 0.7$ between SFG and RQ AGN and 
$\Delta( \log{L_{\rm 1.4 \rm GHz}}) = 0.38 \pm 0.15$ between RQ and RL AGN.
 
As shown in the right hand panel of Figure~\ref{gm_sfg_rq_rl_hist.fig} and Table~\ref{tab:sfg_rq_rl_med}, the median $L_{\rm IR}$ and the width of the distributions increase from SFG, to
RQ AGN and RL AGN. 
Figure~\ref{gm_sfg_rq_rl_hist.fig} indicates the $L_{\rm IR}$ ranges corresponding to NIRGs, LIRGs, ULIRGs 
and HyLIRGs (see \cite{2013MNRAS.428..291P}), and Table~\ref{sfg_rq_rl_IR} shows the breakdown of
the number of sources within each range for each class. 
The majority of the SFGs have $L_{\rm IR}$ in the range of LIRGS and ULIRGS.
RQ AGNs are evenly distributed over LIRGs, ULIRGs and HyLIRGs, with very few in the NIRGS range.
RL AGNs are more broadly distributed with the dominant faction at the low $L_{\rm IR}$ range.

\begin{table*}
\caption{Number of NIRGs, LIRGs, ULIRGs and HyLIRGs in the subsample according to their star-forming, radio-quiet 
and radio-loud FIR luminosities.}
\centering
\begin{tabular}{clccc}
\hline
\hline
$\log_{10}(L_{\rm IR}/L_{\odot})$\,range  &Type  & SFG\# & RQ\,AGN\# & RL\,AGN\#\\
\hline  
$\leq11$            &  NIRG    & 149    & 11 & 23 \\
$11 - 12$   	&  LIRG    & 439    & 45 & 14 \\
$12\ - 13$  	& ULIRG    & 339    & 48 &  8 \\
$\geq13$         &HyLIRG    & 155    & 45 & 10 \\
\hline
\end{tabular}
\label{sfg_rq_rl_IR} 
\end{table*}

\subsection{The Far-Infrared Radio Correlation}\label{The Far-Infrared Radio Correlation}

\begin{figure*}
\includegraphics[width=0.8\textwidth]{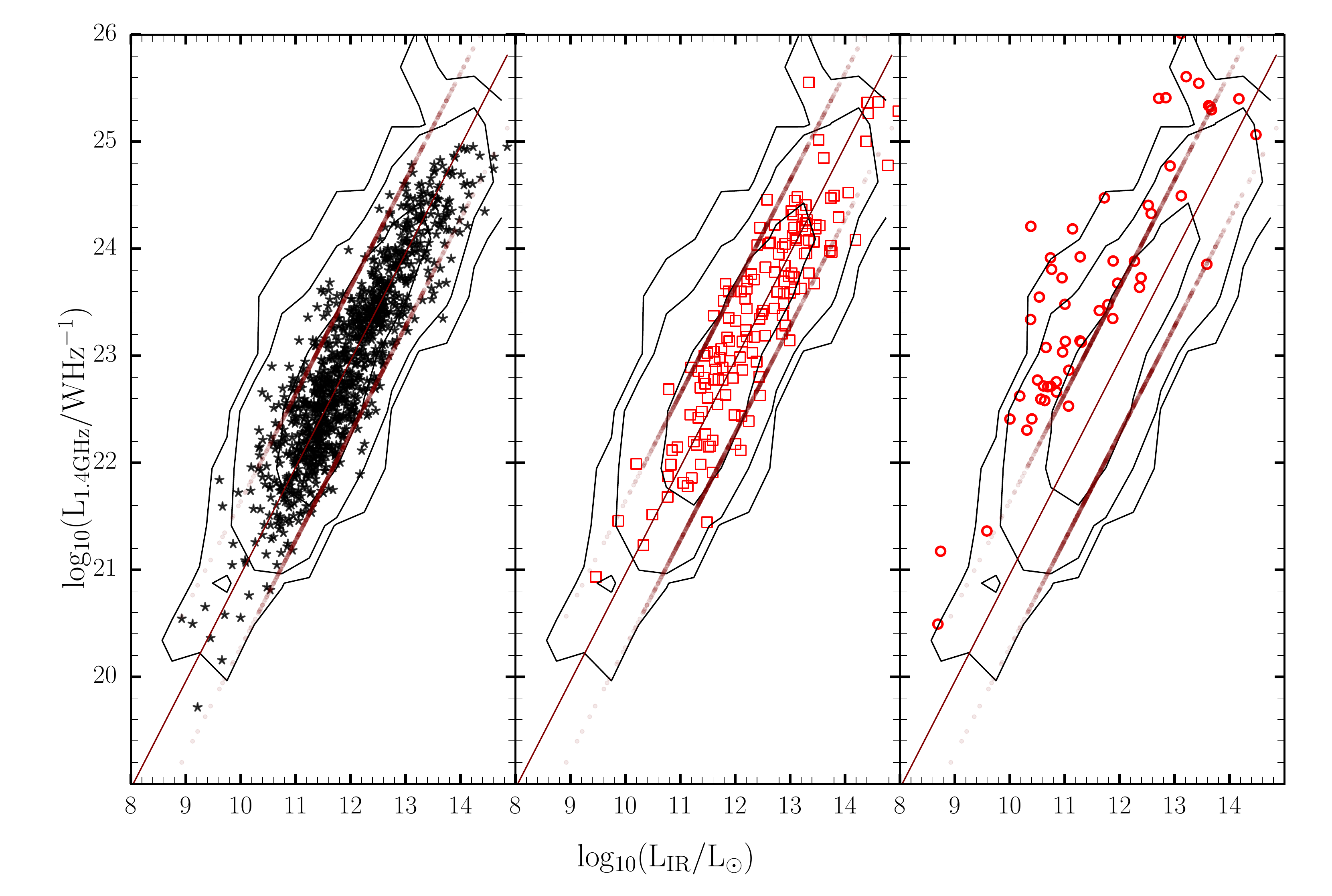}
\caption{$\log_{10}(L_{\rm 1.4\,GHz})$ versus $\log_{10}(L_{\rm IR}/L_{\odot})$ for SFG (first panel), RQ AGN 
(middle panel) and RL AGN (last panel) according to the AGN selection described in Section~\ref{diagnostics.sec}. 
The contours show 1$\sigma$, 2$\sigma$ and 3$\sigma$ confidence intervals of the GMRT sources with radio and IR luminosity respectively. The diagonal solid line in each panel show  the $\rm{q_{IR}}$ relation given in equation~\ref{qIR} for the SFGs.} 
\label{gm_sfgs_agns_contour_new}
\end{figure*}
   
We characterised the IR/radio properties of our sources by the logarithmic ratio between the IR bolometric
 (8-1000 $\mu$m) luminosity and the Radio 1.4 GHz luminosity $q_{\rm{IR}}$\citep{1985ApJ...298L...7H}:
\begin{equation}
q_{\rm IR}  = \log_{10} \left( \frac {L_{\rm IR}} {3.75 \times 10^{12}\,{\rm W} } \right ) - \log_{10} \left ( \frac{ L_{\rm 1.4\,GHz}} {\rm W\,Hz^{-1}} \right)
\label{qIR}
\end{equation}
Figure~\ref{gm_sfgs_agns_contour_new} presents radio power at 1.4 GHz versus FIR luminosity for SFGs detected 
by our classification. The contours represent the 1$\sigma$, 2$\sigma$ and 3$\sigma$ of the GMRT sources with radio and IR luminosity. 
The diagonal solid line in each panel show  the $\rm{q_{IR}}$ relation given in equation~\ref{qIR} for the SFGs 
since the FIRC is believed to be driven mostly by star formation \citep{1992ARA&A..30..575C, 2001ApJ...554..803Y}. 
The dotted lines in all the panels indicate the $\pm$ 1$\sigma$ dispersion in the value of the calculated $q_{\rm IR}$. 
The SFGs have a tight FIRC, with a  median $q_{\rm IR}$ value of $2.45 \pm 0.01$. The selected RQ AGNs also show 
a tight correlation in agreement with the FIRC, with a median $q_{\rm IR}$ value of $2.47 \pm 0.04$. 
RL AGNs lie well above the FIRC for SFGs and RQ AGNs with a median $q_{\rm IR}$ value of $1.43 \pm 0.07$. 
This difference for RL AGNs can be attributed to the additional AGN component to radio emission 
\citep{2015MNRAS.452.4111R}. 

Our median ${q_{IR} = 2.45 \pm 0.01}$ for SFG is in good agreement with a measurement,
e.g. \cite{2010A&A...518L..31I}), who used submillimetre galaxy (SMG) samples with radio emission to
measure a median $q_{\rm IR} = 2.40 \pm 0.2$ for a sample of IR-bright ($S_{250 \mu \rm m} \gtrsim 20$\,mJy) 
galaxies out to $z  <  3.5$. 
Our median $q_{\rm IR}$ is higher, however, than the value of $q_{\rm IR} = 2.17 \pm 0.1$, as measured by 
\cite{2010A&A...518L..28M} for a sample of bright  $(S_{850 \mu \rm m}  >  5$\,mJy) single-dish submillimetre sources
observed in GOODS-N, and lower than that reported by \cite{2014MNRAS.442..577T}, who measured a median $q_{\rm IR} = 2.56 \pm 0.05$ for 870 $\mu{\rm m}$-selected submillimetre galaxies, observed at high resolution 
with ALMA in the Extended Chandra Deep Field South. Our measurement of $q_{\rm IR}$ is also lower than  that  
reported by \cite{2003ApJ...586..794B} who measured $q_{\rm IR} = 2.64 \pm 0.02$ for 162 galaxies with IR and
radio data and no signs of AGN by assembling a sample of SFGs with FUV, IR and radio data to quantitatively 
explore the radio-IR correlation.
Our value of $q_{\rm IR}$ from a larger low-frequency selected sample lies thus within the range of variation of values derived from submillimetre selected samples.

\subsection{Radio emission in SFGs and RQ AGNs}\label{sfgs-and-rq-agns.sec}
\begin{figure}
  \includegraphics[width = \columnwidth]{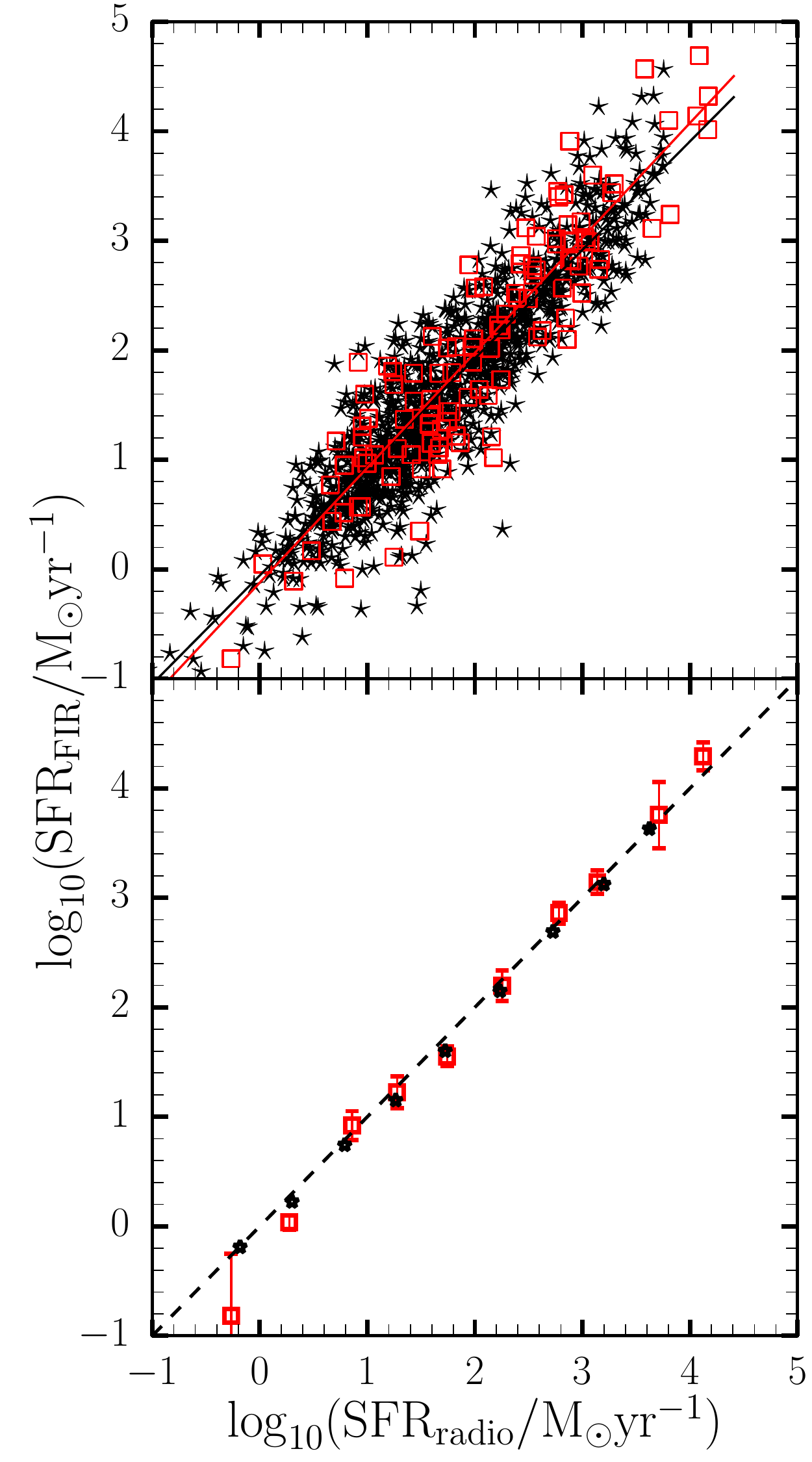}
  \caption{SFR derived from the FIR luminosity versus the SFR from the radio luminosity. SFGs are plotted as black 
  stars and RQ AGNs as open red squares. The bottom panel show the binned $\log_{10}({\rm SFR})$ for the SFGs and the RQ AGNs in bin width of 0.5$\log_{10}({\rm SFR})$} 
  \label{gm_sfr.fig} 
\end{figure}

\cite{1993MNRAS.263..425M} suggested that RQ AGNs are a scaled-down version of RL AGNs at lower 
radio power whilst \cite{1991MNRAS.251..112S} found that the major contribution to the radio emission in 
these system is due to the star formation in the host galaxy. However, the study of the cosmological evolution 
and luminosity function by \cite{2011ApJ...740...20P}, found that their radio emissions are significantly different for the two types of AGNs, while they are indistinguishable for SFGs and RQ AGNs. We characterize nature of the radio
emission from the three types by comparing  the star formation rates inferred from the IR and radio. 
The empirical conversion between the radio power at 1.4 GHz and the SFR of the galaxy according to 
\citep{2011ApJ...737...67M} is:
\begin{equation}
{\rm SFR_{radio}} [M_{\odot}\,{\rm yr}^{-1}] = \log_{10}(L_{\rm 1.4\,GHz}) - 21.20
\label{sfr_rad}
\end{equation}
 In Figure~\ref{gm_sfr.fig} we compare the SFR computed from the FIR luminosities by \cite{RowanRobinson2013} 
 with the SFR derived from the radio luminosities for our classified SFGs and RQ AGN.  A regression fit to the data yields: 
\begin{multline}
\log_{10}({\rm SFR_{FIR}}) = 0.95\pm0.02\times\log_{10}({\rm SFR_{radio}})
\mathrm{-0.02\pm0.03}
\label{sfr_sfg1}
\end{multline}
for all the sources with redshifts having infrared and radio luminosities, while for SFG alone we obtain
\begin{multline}
\log_{10}({\rm SFR_{FIR}})_{\rm SFG} = 
0.98\pm0.02\times\log_{10}({\rm SFR_{radio}})_{\rm SFG}
\\
\mathrm{- 0.07\pm0.03}
\label{sfr_sfg}
\end{multline}
and for RQ AGN
\begin{multline}
\log_{10}({\rm SFR_{FIR}})_{\rm RQ\,AGN} = 
1.04\pm0.05\times\log_{10}({\rm SFR_{radio}})_{\rm RQ\,AGN}
\\
\mathrm{- 0.12\pm0.11}
\label{sfr_rq_agn}
\end{multline}
The fits for the SFGs and RQ AGNs (shown as black and red lines respectively in Figure~\ref{gm_sfr.fig}), 
are not significantly different within errors, suggesting that the main contribution to the radio emission in RQ AGN
is due to the star formation in the host galaxy rather than black hole activity. 
Thus radio power is a good tracer for the SFR also in RQ AGNs. 

\subsection{Star forming galaxies at high redshift}

\begin{figure}
\includegraphics[width = \columnwidth]{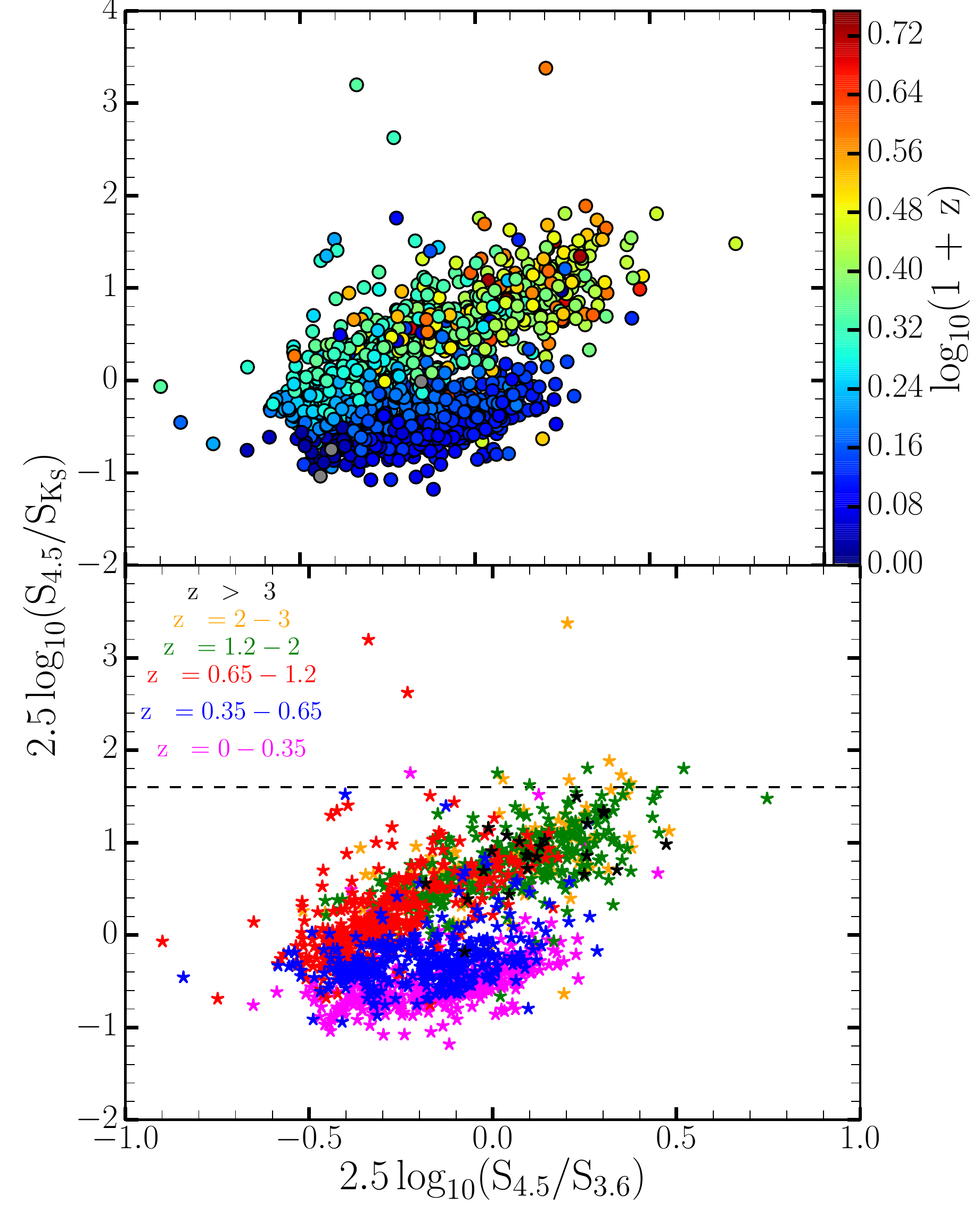}
\caption{K$\rm{_{s}}$, 3.6 and 4.5 color-color diagram. Top panel: SFGs color coded with redshift. 
Bottom panel: SFG sources with same redshift demarcation ranges following \citep{2012ApJ...744..155W}.}
\label{gm_kmag_sfg.fig} 
\end{figure}

The SFG in our radio sample, while predominantly at z < 2, include some objects with photometric redshifts
much larger, including 27 sources with $z\,\geq\,3$. 
We investigate the reliability of such high redshifts by following the analysis of \cite{2012ApJ...744..155W}, 
who presented a new color selection of extremely red objects (EROs) (i.e. most extreme dust-hidden high-redshift galaxies) with ${K_{S}}$ and IRAC colors of $K_{S} - 4.5\mu {\rm m} > 1.6$. The photometric redshifts used by \cite{2012ApJ...744..155W} to select EROs are between 1.5 and 5, with $\sim$70$\%$ at $z\sim 2 - 4$.
This  selection  aims  at  galaxies  at $z > 2$  whose extremely red colors are likely caused by large dust extinction.
We use their approach to examine our SFGs sample and in particular where our objects at $z > 3$ reside in the diagnostic plot.
Figure~\ref{gm_kmag_sfg.fig} presents the  $K_{S}$, 3.6 and 4.5 color-color diagram for our selected SFGs color 
coded according to redshift (top panel) and, also the same sample with the same redshift demarcations used by 
\cite{2012ApJ...744..155W}. 
The top panel shows that SFGs at high redshifts have $2.5\log_{10}(S_{4.5}/S_{K_{S}}) > 0$,
consistent with this high redshift selection criterion.

\section{Conclusion}\label{sum.sec}

This work was aimed at disentangling the different radio populations that contribute to the faint radio sky. 
We have explored different AGN indicators from the radio, mid-infrared, optical and X-ray. We reaffirm that the 
ratio between the mid-infrared and radio flux parametrised by the $q_{24\mu m}$ value, demonstrated in 
\cite{2013MNRAS.436.3759B}, is an important parameter to identify RL AGNs. Our scheme expands on the one adopted by \cite{2013MNRAS.436.3759B} and combines radio, mid-infrared, optical and X-ray data to efficiently separate the radio source population with redshift associations into three classes: SFGs, RQ AGNs and RL AGNs.

We have determined the relative contribution of the three classes of sources to the subsample of 1526 radio sources with redshifts and at least one multi-wavelength diagnostic and find that $\sim$80\% are SFGs, $\sim$12\% are RQ AGNs and $\sim$8\% are RL AGNs. 
Compared to previous analysis of sources over smaller areas at 1.4 GHz in the ECDFS, our result indicates a continued increase in the relative fraction of SFG with decreasing flux density, and confirm that RQ objects dominate the AGN population.
The significantly higher fraction of SFG in our sample may also partially arise from the selection at lower frequency, where at a given flux density threshold flat-spectrum AGN cores are preferentially detected at 1.4 GHz. 
Multi-frequency investigations to explore the spectral index distribution of $\mu$Jy radio sources will help resolve this question.

The median redshift of our radio sources for all three categories is $\sim$1, although in all cases detections span the entire range up to $z > 3$. The median values of both the radio and infrared luminosity systematically increases from SFG, to RQ AGN and RL AGN (see Table~\ref{tab:sfg_rq_rl_med}).  
 
The  median $q_{\rm 24 \mu m}$ for SFG, $0.89\pm0.01$ is slightly below that for RQ AGN, $1.05\pm0.03$, but both
differ substantially from the value for RL AGN of $-0.06\pm0.07$.
In contrast to $q_{\rm 24 \mu m}$ , SFG and RQ AGN show no difference in the ratio of radio to bolometric
IR luminosity, with $q_{\rm IR}$  of  of $2.45 \pm 0.01$ and $2.47 \pm 0.04$ respectively.
The Far-IR Radio correlation ($q_{\rm IR}$) seen in the SFGs is in good agreement with the value found locally 
\citep{2002ApJ...568...88Y} and at higher redshift \citep{2010A&A...518L..31I, 2011MNRAS.413.1777S}.
Although the median radio luminosity of the RQ AGN population is slightly higher than the SFGs, there is no distinction between the objects in Far-IR radio correlation, suggesting that the radio emission from RQ AGN host galaxies results primarily from star formation activity. 
The RL AGN on the other hand are systematically above the Far-IR correlation for SFG and RQ AGN, with a median $q_{\rm IR}$ of $1.54\pm 0.06$, indicating the presence of additional AGN-powered radio emission. 

The need to better understand the faint radio source population in the high redshift universe is one of the 
key motivation for the construction  of the next generation of very large radio telescopes such as the SKA.
In this study we presented an multi-wavelength analysis to classify the $\mu$Jy population.
It is clear that SFG will form a high and growing fraction of the radio sources as we probe deeper.  
Next-generation surveys will thus probe beyond the AGN phenomenon to the underlying radio astrophysics of
the evolution of galaxies and star formation.
The GMRT data presented here in total intensity also includes full-Stokes polarimetry.  An exploration of the 
polarisation properties of this faint radio source population will be the subject of a subsequent paper.

\section*{Acknowledgements}
The authors acknowledge support from the Square Kilometre Array South Africa project, the South African National Research Foundation and Department of Science and Technology.  E.F.O. acknowledges funding from the National Astrophysics and Space Science Programme.
MV acknowledges support from the European Commission Research Executive Agency (FP7-SPACE-2013-1 GA 607254), the South African Department of Science and Technology
(DST/CON 0134/2014) and the Italian Ministry for Foreign Affairs and International Cooperation (PGR GA ZA14GR02).

We thank the staff of the GMRT that made these observations possible. GMRT is run by the National Centre for Radio Astrophysics of the Tata Institute of Fundamental Research.

% Funding for SDSS-III has been provided by the Alfred P. Sloan Foundation, the Participating Institutions, the National Science Foundation, and the U.S. Department of Energy Office of Science. 
% The SDSS-III web site is http://www.sdss3.org/. SDSS-III is managed by the Astrophysical 
% Research Consortium for the Participating Institutions of the SDSS-III Collaboration including 
% the University of Arizona, the Brazilian Participation Group, Brookhaven National Laboratory, 
% Carnegie Mellon University, University of Florida, the French Participation Group, the German 
% Participation Group, Harvard University, the Instituto de Astrofisica de Canarias, the 
% Michigan State/Notre Dame/JINA Participation Group, Johns Hopkins University, 
% Lawrence Berkeley National Laboratory, Max Planck Institute for Astrophysics, 
% Max Planck Institute for Extraterrestrial Physics, New Mexico State University, 
% New York University, Ohio State University, Pennsylvania State University, University of Portsmouth,
%  Princeton University, the Spanish Participation Group, University of Tokyo, University of Utah,
%   Vanderbilt University, University of Virginia, University of Washington, and Yale University.

This work is based in part on observations made with the \textit{Spitzer Space Telescope}, which is operated by the Jet Propulsion Laboratory, California Institute of Technology under a contract with NASA.

\bibliographystyle{mnras}
\bibliography{main} 

% Don't change these lines
\bsp	% typesetting comment
\label{lastpage}
\end{document}